\documentclass{aa}

\usepackage[varg]{txfonts}
\usepackage{graphicx}
\usepackage{natbib}
\usepackage{multirow}
\usepackage{subcaption}
\usepackage{stfloats}
\usepackage[colorlinks = true, linkcolor = blue, citecolor = blue]{hyperref}
\bibpunct{(}{)}{;}{a}{}{,} 
\usepackage{amstext}

\begin{document}

\title{HD142527: Quantitative disk polarimetry with SPHERE\thanks{Based on observations collected at the European Southern Observatory under ESO programmes 095.C-0273(A), 096.C-0248(B), 098.C-0790(A), 099.C-0127(B) and 099.C-0601(A)}}

\subtitle{}

\author{S.~Hunziker\inst{\ref{instch1}} 
   \and H.M.~Schmid\inst{\ref{instch1}}
   \and J.~Ma\inst{\ref{instch1}}
   \and F.~Menard\inst{\ref{instf1}}    
   \and H.~Avenhaus\inst{\ref{instch1}}
   \and A.~Boccaletti\inst{\ref{instf4}}      
   \and J.L.~Beuzit\inst{\ref{instf3}}
   \and G.~Chauvin\inst{\ref{instf1}}   
   \and K.~Dohlen\inst{\ref{instf3}}     
   \and C.~Dominik\inst{\ref{instnl2}}
   \and N.~Engler\inst{\ref{instch1}}     
   \and C.~Ginski\inst{\ref{instnl2}}  
   \and R.~Gratton\inst{\ref{insti1}}
   \and T.~Henning\inst{\ref{instd1}}      
   \and M.~Langlois\inst{\ref{instf7},\ref{instf3}}      
   \and J.~Milli\inst{\ref{instf1}}         
   \and D.~Mouillet\inst{\ref{instf1},\ref{instf2}} 
   \and C.~Tschudi\inst{\ref{instch1}} 
   \and R.G.~van Holstein\inst{\ref{instnl3}}
   \and A.~Vigan\inst{\ref{instf3}}}


\institute{
ETH Zurich, Institute for Astronomy, Wolfgang-Pauli-Strasse 27, 
CH-8093 Zurich, Switzerland\\
\email{silvan.hunziker@phys.ethz.ch}\label{instch1}
\and
LESIA, CNRS, Observatoire de Paris, Universit\'{e} Paris Diderot, UPMC, 5 place J. Janssen, 92190 Meudon, France\label{instf4}
\and
Aix Marseille Universit\'{e}, CNRS, LAM (Laboratoire d’Astrophysique de Marseille) UMR 7326, 13388, Marseille, France\label{instf3}
\and
Universit\'{e} Grenoble Alpes, IPAG, 38000 Grenoble, France\label{instf1}
\and
Anton Pannekoek Astronomical Institute, University of Amsterdam, PO Box 94249, 1090 GE Amsterdam, The Netherlands\label{instnl2}
\and
INAF – Osservatorio Astronomico di Padova, Vicolo dell’Osservatorio 5, 35122 Padova, Italy\label{insti1}
\and
Max-Planck-Institut f\"{u}r Astronomie, K\"{o}nigstuhl 17, 69117 Heidelberg, Germany\label{instd1}
\and
Centre de Recherche Astrophysique de Lyon, CNRS/ENSL Universit\'{e} Lyon 1, 9 av. Ch. Andr\'{e}, 69561 Saint-Genis-Laval, France\label{instf7}
\and
CNRS, IPAG, 38000 Grenoble, France\label{instf2}
\and
Leiden Observatory, Leiden University, P.O. Box 9513, 2300 RA
Leiden, The Netherlands\label{instnl3}}

\date{Received --- ; accepted --- }

\abstract{}
{We present high-precision photometry and polarimetry based on visual and near-infrared imaging data for the protoplanetary disk surrounding the Herbig Ae/Be star HD142527, with a strong focus on determining the light scattering parameters of the dust located at the surface of the large outer disk.}
{We re-reduced existing polarimetric differential imaging data of HD142527 in the VBB (735~nm) and $H$-band (1625~nm) from the ZIMPOL and IRDIS subinstruments of SPHERE at the VLT. With polarimetry and photometry based on reference star differential imaging (RDI), we were able to measure the linearly polarized intensity and the total intensity of the light scattered by the circumstellar disk with high precision. We used simple Monte Carlo simulations of multiple light scattering by the disk surface to derive constraints for three scattering parameters of the dust: the maximum polarization of the scattered light $P_{\rm max}$, the asymmetry parameter $g$, and the single-scattering albedo $\omega$.}
{We measure a reflected total intensity of $51.4\pm1.5$~mJy and $206\pm12$~mJy and a polarized intensity of $11.3\pm0.3$~mJy and $55.1\pm3.3$~mJy in the VBB and $H$-band, respectively. We also find in the visual range a degree of polarization that varies between $28\%$ on the far side of the disk and $17\%$ on the near side. In the $H$-band, the degree of polarization is consistently higher by about a factor of 1.2. The disk also shows a red color for the scattered light intensity and the polarized intensity, which are about twice as high in the near-infrared when compared to the visual. We determine with model calculations the scattering properties of the dust particles and find evidence for strong forward scattering ($g\approx 0.5-0.75$), relatively low single-scattering albedo ($\omega \approx 0.2-0.5$), and high maximum polarization ($P_{\rm max} \approx 0.5-0.75$) at the surface on the far side of the disk for both observed wavelengths. The optical parameters indicate the presence of large aggregate dust particles, which are necessary to explain the high maximum polarization, the strong forward-scattering nature of the dust, and the observed red disk color.}
{}


\authorrunning{Hunziker et al.}
\maketitle

\section{Introduction}
The existence of circumstellar disks was first deduced from infrared (IR) emission components in unresolved observations of the spectral energy distribution (SED) of young stars. Circumstellar disks reflect some of the stellar light at visual and near-IR wavelengths, and some of the light is absorbed by the disk and re-emitted as excess in the mid- and far-IR. The disk SED contains important information about the disk structure and the dust properties, and IR emission bands can be used to determine the composition of the dust. The clear detection of reflected light from protoplanetary disks was initially achieved with the \textit{Hubble} Space Telescope (HST) and with adaptive optics (AO) assisted, ground-based telescopes \citep{Roddier95, Silber00}. More recently, the development of dedicated instruments with polarimetric differential imaging (PDI) capabilities such as the Spectro-Polarimetric High-contrast Exoplanet REsearch instrument (SPHERE; \citet{Beuzit19}) at the Very Large Telescope (VLT), the Gemini Planet Imager (GPI; \citet{Perrin15}) at the Gemini South Telescope and the High-Contrast Coronographic Imager for Adaptive Optics (HiCIAO; \citet{Hodapp08}) at the Subaru Telescope have opened up the possibility to perform observations of reflected light from circumstellar disks with unprecedented sensitivity. PDI has proven to be an effective and reliable method to remove unpolarized stellar light and reveal the polarized reflected light of disks. These observations have led to the detection of many protoplanetary disks in scattered light and revealed disks with a surprisingly large variety of different masses, sizes, and morphologies \citep[e.g.,][]{Takami14, Garufi17, Monnier17, Avenhaus18}. 

Protoplanetary disks are considered to be the places where planet formation occurs around young pre-main-sequence stars. The disks initially consist of a mixture of gas and submicron-sized dust particles. Dust coagulation over time leads to the growth of dust particles sizes and eventually to the formation of planetesimals and planets. The analysis of the dust constituents and dust evolution mechanisms in young stellar environments is therefore vital for our understanding of planet formation. One way of constraining the dust properties is through the optical properties of the reflected light. A few studies in the past have used direct imaging of circumstellar disks to determine optical properties such as the scattering asymmetry parameter $g$, the maximum polarization of the reflected light $P_{\rm max}$, the single-scattering albedo $\omega,$ and the (polarized) scattering phase function \citep[e.g.,][]{Duchene04, Pinte08, Mulders13}. These studies have shown, for example, that the wavelength dependence of the scattered light intensity for disks can significantly deviate from measurements performed on the smaller sized dust found in the interstellar medium or molecular dust clouds \citep[e.g.,][]{Duchene04}, which is an indication for dust growth. The measurements in circumstellar disks also revealed degrees of linear polarization that are not compatible with scattering on simple spherical and compact dust grains \citep[e.g.,][]{Pinte08}.

Transition disks are well suited for scattered-light observations because they usually feature a dust-depleted inner cavity between the star and the inner disk wall. This results in a bright illuminated inner disk wall at a large enough distance from the star so that it can be observed in nearby star-forming regions. HD142527 has a very large inner cavity, most likely because the central star has a companion. Therefore it is not a typical transition disk \citep{Price18}, but its brightness and size make it exceptionally well suited for scattered-light observations.

In this paper, we focus on the quantitative analysis of high spatial resolution visual and near-IR observation of the HD142527 protoplanetary disk to determine the dust optical properties. The data were taken with the PDI modes of the Zurich IMaging POLarimeter \citep[ZIMPOL;][]{Schmid18} and the InfraRed Dual-band Imager and Spectrograph \citep[IRDIS;][]{Dohlen08, vanHolstein20, deBoer20}, which are subsystems of SPHERE. This allows the simultaneous measurement of the total intensity and linearly polarized intensity of the scattered light and therefore also determine the degree of polarization for different parts of the HD142527 disk with high precision. For simplicity, we refer from here on to linear polarization just as polarization. The scattered intensities from the near and far side of the disk constrain the asymmetry parameter $g$ of the scattering phase function, the degree of polarization, and the single-scattering albedo $\omega$, while $90^{\circ}$ scattering constrains the maximum polarization $p_{\rm max}$.

We provide accurate photopolarimetry of the reflected light as a function of the azimuthal angle and radial distance for the large outer disk around HD142527 and set limits on scattering parameters that can be compared to measurements from other disks, to model simulations, or laboratory studies of dust light-scattering properties. The outer disk is hereafter called the disk, and whenever the inner disk or the hot dust are discussed, we specify this accordingly. In Sec. 2 we review properties and previous observations of HD142527 relevant for this work. In Sec. 3 we describe the observations and data reduction, and in Sec. 4 we describe the data analysis procedures used to measure the scattered intensity, the polarized intensity, and the degree of polarization of the reflected light on different parts of the disk. In Sec. 5 we present the main results of our measurements, which are discussed in Sec. 6, and Sec. 7 contains our concluding remarks. 

\section{Dust properties in HD142527}
HD142527 is a binary system that consists of a massive F-type Herbig Ae/Be star with an M-dwarf companion \citep{Biller12} on a close-in ($<0.1\arcsec$) and possibly highly eccentric orbit \citep{Claudi19}. The system has been studied intensively in the past due to its extended disk at a distance of only 156~pc \citep{GAIA18} and its unusually high IR excess of $F_{IR}/F_* = 0.92$ \citep{Dominik03}. The inner rim of the large circumbinary disk is slightly eccentric ($\epsilon\approx0.137$), with a semimajor axis of about 140~AU \citep{Avenhaus14}, and the disk is optically thick in the visual and the near- and mid-IR. \citet{Fukagawa06} have detected reflected light from the disk at separations up to around 550~AU in $H$-band observations. The structure of the disk is still investigated because the system is complex and the outer disk is partially shadowed by a compact hot disk near the star \citep{Marino15}. \citet{Verhoeff11} have found that the high IR excess of the system can be explained by invoking the presence of a relatively high and steep inner wall for the disk at 130~AU, but no clear consensus exists on the scale height of the disk and its inclination toward the observer because different methods find slightly different solutions. Radial velocity measurements of molecular lines with ALMA found a disk inclination of about $28^{\circ}$ with a position angle of $-20^{\circ}$ \citep{Fukagawa13, Perez15}, but the results depend on the poorly constrained stellar mass of HD142527, which was estimated to be around 2~M$_{\odot}$. Lower inclinations around $20^{\circ}-24^{\circ}$ were found by fitting IR imaging data \citep{Verhoeff11, Avenhaus14}, but these results depend on assumptions about the light-scattering model and can be strongly correlated with the scale height of the disk \citep{Avenhaus14}. The values for the scale height of the inner disk wall were also derived by fitting scattered-light images and the IR SED, and the results vary between 17 and 30~AU \citep{Verhoeff11, Avenhaus14, Min16}. Accurate geometric parameters for the disk are important for interpreting the scattered intensity and degree of polarization because these signals strongly depend on the scattering angle.

The $H-K$ color of the disk was determined to be gray \citep{Fukagawa06},
and \citet{Honda09} revealed a strong water-ice absorption feature at 3.1$\mu$m. The presence of water ice in this system was previously detected by a mineralogical analysis of the HD142527 spectrum presented in \citet{Malfait99}. Water ice is important from the planet formation perspective because ice-coated dust grains are expected to stick together more easily and form larger grains more efficiently.

Because of the large size and brightness of the disk around HD142527, the emission from the dust has been imaged extensively at wavelengths ranging from around 0.6~$\mu$m \citep{Avenhaus17} up to 24~$\mu$m \citep{Fujiwara06} and at submillimeter (submm) and radio wavelengths up to 1.2~mm \citep{Perez15}. The gas distribution and dynamics have been studied with submm observations of emission lines \citep[e.g.,][]{Casassus13, Casassus15, Fukagawa13, Rosenfeld14, Perez15}. In observations of the scattered-light intensity in the near-IR, the near side of the disk is brighter than the far side \citep[e.g.,][]{Fukagawa06, Avenhaus14} because the dust scatters light predominantly in forward direction. For mid-IR observations at $\lambda > 10~\mu$m, the far side of the disk is brighter than the near side because these observations image the thermal emission from the strongly illuminated inner wall of disk, which is exposed on the far side but hidden from view on the near side.

\citet{Canovas13} and \citet{Avenhaus14} both used $H$- and $K$-band polarimetry of HD142527 to determine the degree of polarization of the light scattered from the disk and found that the far side produces significantly more linear polarization than the near side. \citet{Min16} used their detailed model of the disk to produce a synthetic scattered-light image, which also reproduces the asymmetry in degree of polarization. However, the values for the degree of polarization from all studies are in poor agreement. With similar observations in $H$- and $K$-band, \citet{Canovas13} measured 15-20\%, while \citet{Avenhaus14} obtained 40-50\%, and \citet{Min16} predicted  a polarization of 50-60\% with their model for the same
wavelengths. In this work, we carefully analyze polarimetric imaging data from well-calibrated instruments to deliver improved measurements of reflected-light properties such as the degree of linear polarization in order to solve these disagreements.

\section{Observations and data reduction}
\begin{table*}
\caption{Summary of all SPHERE/ZIMPOL and IRDIS observations that were used in this work. For each dataset we list the observing conditions (seeing in arcseconds and coherence time $\tau_0$ in ms) and the wind speeds during the observations.}
\label{table: data}
\centering
\begin{tabular}{l l l l l l l l l l l l}
\hline\hline
Date (UT) & Object & m$_{\rm V}$ & Filter & NDIT $\times$ DIT & \# of & $t_{\rm exp}$\tablefootmark{a} & Seeing & $\tau_0$ & Air mass & Wind\\
 & & [mag] & & & pol. & [min] & $\left(\arcsec\right)$ & [ms] & & speed \\
 & & & & & cycles & & & & & [m/s] \\
 
\hline
        2015/05/02 & HD142527 & 8.3 & VBB & 14 $\times$ 3~sec & 21 & 58.8 & 0.5--0.8 & 2.3--3.4 & 1.16--1.42 & 4.0--4.9\\
        2016/03/31 & HD142527 & 8.3 & VBB & 18 $\times$ 2~sec & 24 & 57.6 & 0.6--1.0 & 3.2--4.8 & 1.19--1.43 & 4.1--5.4\\
        2017/05/31 & HD142527 & 8.3 & VBB & 20 $\times$ 3~sec & 17 & 68 & 0.6--0.8 & 7.5--13 & 1.05--1.09  & 1.7--2.4\\
    2017/06/19 & $\alpha$~Cen~B\tablefootmark{b} & 1.3 & VBB & 20 $\times$ 1.1~sec & 141 & 206.8 & 0.3--1.0 & 4.5--9.5 & 1.24--1.52 & 1.0--6.0\\
\hline
    2016/10/11 & HR8799\tablefootmark{b} & 6 & BB\_H & 3 $\times$ 16~sec & 43 & 137.6 & 0.4--0.9 & 2.4-6.0 & 1.43--1.66 & 2.2--5.0\\
    2017/05/31 & HD142527 & 8.3 & BB\_H & 8 $\times$ 16~sec & 3 & 25.6 & 0.6--0.8 & 10--12 & 1.05--1.13 & 0.5--2.2\\
    2017/05/31 & HD142527 & 8.3 & BB\_H & 12 $\times$ 8~sec & 6 & 38.4 & 0.6--0.9 & 10--15 & 1.05--1.09 & 0.0--1.5\\
\hline
\end{tabular}
\tablefoot{ 
\tablefoottext{a}{The total exposure time per channel}
\tablefoottext{b}{Observations used to model the reference surface brightness profiles without the disk signal}
}
\end{table*}

\begin{figure}
\centering
\includegraphics[width=\linewidth]{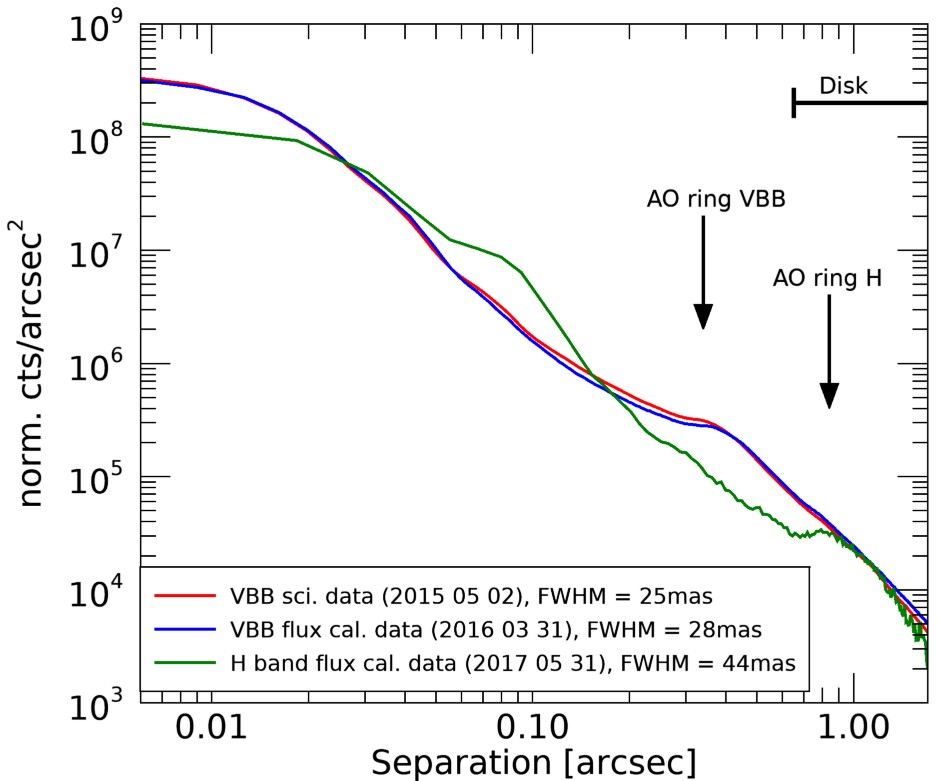}
\caption{Normalized PSFs of HD142527 for the ZIMPOL VBB and IRDIS $H$-band normalized to a total count level of $10^6$ within an aperture diameter of 3~arcsec. The bump at $\sim0.08\arcsec$ in the $H$-band PSF is caused by the low-wind effect. The locations of the AO control rings and the disk are indicated.}
  \label{fig:data_prof_reso}
\end{figure}

\subsection{ZIMPOL observations}
\label{sec:ZIMPOL observations}
The main ZIMPOL data we used to measure the disk signal in this work were obtained on 2015 May 2 as part of the SPHERE GTO program and were first published in \citet{Avenhaus17}. The observations were obtained in P2 polarimetry mode (field-stabilized) with four different derotator offset positions to reduce the fixed pattern noise induced by the instrument. The images were taken in the very broadband (VBB) filter that spans over the $R$ and $I$-band ($\lambda_c = 735~{\rm nm}, \Delta\lambda = 290~{\rm nm}$) in both channels of ZIMPOL with 3~sec detector integration time (DIT), and without the coronagraph. The high throughput of the VBB filter is well suited to detect the faint signal of the reflected light in the visible. The observations were obtained with the detectors in fast-polarimetry mode to reduce the residual speckle noise in the differential polarization at small separations because the original goal of the observations was the detection of the inner disk. The data were minimally saturated in the core of the point spread function (PSF) with the 3~sec DIT. We used these data to measure the scattered disk intensity and polarized intensity because they provide the highest Strehl ratio of all ZIMPOL datasets listed in Table~\ref{table: data}.

A second dataset with ZIMPOL was obtained as part of the GTO program on 2016 March 31 with DIT~=~2sec to avoid saturation. The data are also described in \citet{Avenhaus17}. Half of the 2016 observations were obtained in P2 polarimetry (field-stabilized), and the other half in P1 polarimetry (field-rotating) mode, but they were otherwise identical to the observations made in 2015. We used these data to determine the total flux from the system $I_{\rm total}$ in the VBB filter because this filter show no detector saturation. The total flux measured in these data is $3\pm0.5\%$ higher than that of the 2015 data after correcting for the different airmasses during the observations. The small difference could be due to saturation of the star in the 2015 data.

A third dataset of HD142527 was obtained on 2017 May 31 for an open time proposal. The observations were identical to the first observation in 2015, but they showed the low-wind effect. Slow wind at the VLT can lead to a degradation of the PSF because it splits the single-peak PSF into a PSF with multiple smaller peaks \citep{Sauvage15, Cantalloube19}, which significantly lowers the resolution and therefore the signal-to-noise ratio (S/N) of the disk observations. \citet{Schmid18} showed an image of the ZIMPOL PSF from this run. We included these data in our analysis for measuring the intrinsic polarization of HD142527 because the recoating of the M1 and M3 telescope mirrors in April 2017 significantly lowered the instrument polarization for observations with ZIMPOL, which enables more accurate measurements of the intrinsic polarization of a source.

We used a reference star differential imaging (RDI) approach to measure the scattered intensity. Angular differential imaging (ADI) is not possible for these data because the observations were performed in field-stabilized mode. The reference star observations were used to determine the ZIMPOL VBB PSF without the disk, and they were obtained with polarimetric imaging of $\alpha$~Cen~B on 2017 June 19. The coronagraphic PSF of these observations can be used as a reference for the non-coronagraphic HD142527 data because the disk is located at a large separation $>0.6\arcsec$, where the shape of the PSF is barely affected by the coronagraph. The reference star dataset does not contain any strong signals other than the stellar PSF. It was observed during similar conditions and exhibits a large diversity of different VBB PSFs. We randomly selected 120 frames distributed over the whole $\alpha$~Cen~B observing run to construct our reference star dataset. More detailed information about all ZIMPOL datasets used in this work can be found in Table~\ref{table: data}.

The ZIMPOL data were reduced with the sz-software (SPHERE/ZIMPOL) pipeline developed at ETH Z\"urich specifically for the reduction of ZIMPOL data. The IDL-based software package performs basic data preprocessing, reduction, and calibration steps, which are essentially identical to the ESO Data Reduction and Handling (DRH) software package developed for SPHERE \citep{Pavlov08}. A detailed technical description of ZIMPOL is presented in \cite{Schmid18}. The basic data reduction steps include bias subtraction, flat fielding, correcting the modulation-demodulation efficiency, and centering of the frames. We used a common estimated mean center for all frames with an accuracy better than 1 pixel (3.6~mas). Perfect individual centering at the level of <~1~mas is not required for measurements of the extended signal of the disk. In addition, we measured and corrected the relative beam shift of the two orthogonal polarization states, subtracted the frame transfer smearing, and corrected for the residual telescope polarization and the intrinsic polarization of the star. The last two steps normalize the integrated Stokes parameters of the central object to zero ($Q = 0$ and $U = 0$), which is required for HD142527 because the star exhibits a relatively strong intrinsic or interstellar polarization of $\sim1\%$. The normalization can affect measurements of the polarized intensity \citep[e.g.,][]{Hunziker20}. This possibility is investigated further in Sec.~\ref{sec:Intrinsic stellar polarization}.

The Stokes $Q$ and $U$ images were then transformed into the azimuthal $Q_{\phi}$ and $U_{\phi}$ basis according to the radial polarization $Q_r$ and $U_r$ defined in \citet{Schmid06b}:

\begin{equation}
\begin{aligned}
Q_{\phi} &= -Q_r=-Q \cos\left(2\phi\right) - U \sin\left(2\phi\right) \\
U_{\phi} &= -U_r= Q \sin\left(2\phi\right) - U \cos\left(2\phi\right).
\end{aligned}
\label{equ: azimuth coord trafo}
\end{equation}

For reflected light from a circumstellar disk with low inclination, most of the polarized intensity is contained in $Q_{\phi}$, while $U_{\phi}$ should be zero. Multiple scattering of stellar light in the disk, crosstalk between $Q$ and $U,$ and stellar or uncorrected instrumental polarization can lead to a nonzero $U_{\phi}$ parameter, but for the observations used in this work, the signal in $U_{\phi}$ was determined to be negligible. 

The format of the reduced images is $1024\times1024$ pixels with a pixel size of $3.6\times3.6$~mas. We estimated the resolution with the full width at half maximum (FWHM) of the PSF ($\sim25$~mas) for the data from 2015 and the flux calibration data from 2016. As shown in Fig.~\ref{fig:data_prof_reso}, both VBB datasets have the same resolution, and the radial profiles are not significantly different at any separation because the observing conditions were similar. Fig.~\ref{fig:data_prof_reso} also shows the location of the AO control ring for all datasets. The AO control ring refers to an increased speckle intensity located at a distance of $20 \lambda/D$ from the center of the stellar PSF, which can interfere with the extraction of faint circumstellar signals. Images of the scattered intensity (Stokes I) and the polarized intensity (Stokes $Q_{\phi}$) of the disk for the VBB observations are shown in Fig.~\ref{fig:HD142527_stokes_I_Qphi}.

\subsection{IRDIS observations}

\begin{figure*}
\centering
\begin{tabular}{cc}
\includegraphics[totalheight=3.3in]{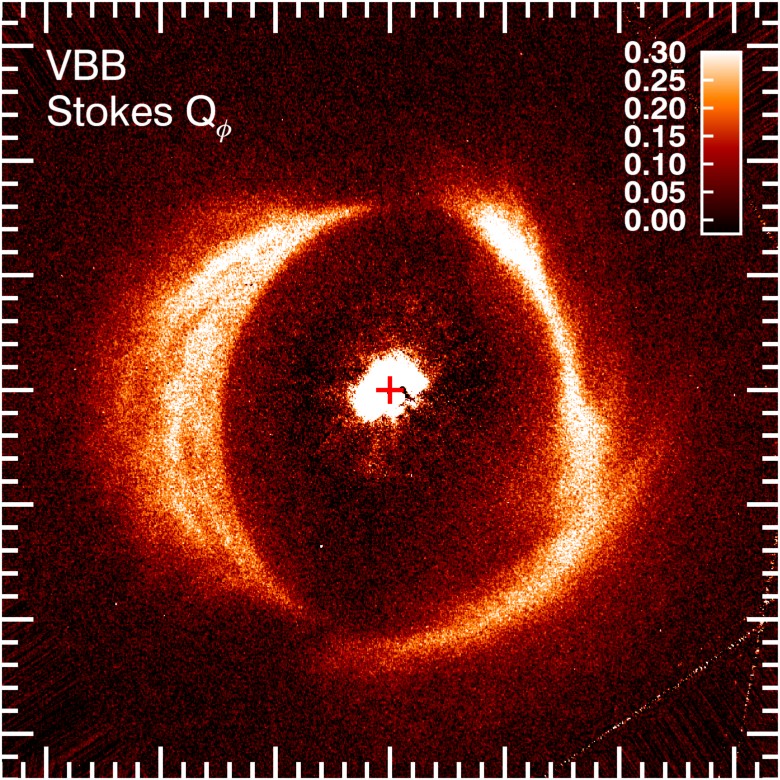} & \includegraphics[totalheight=3.3in]{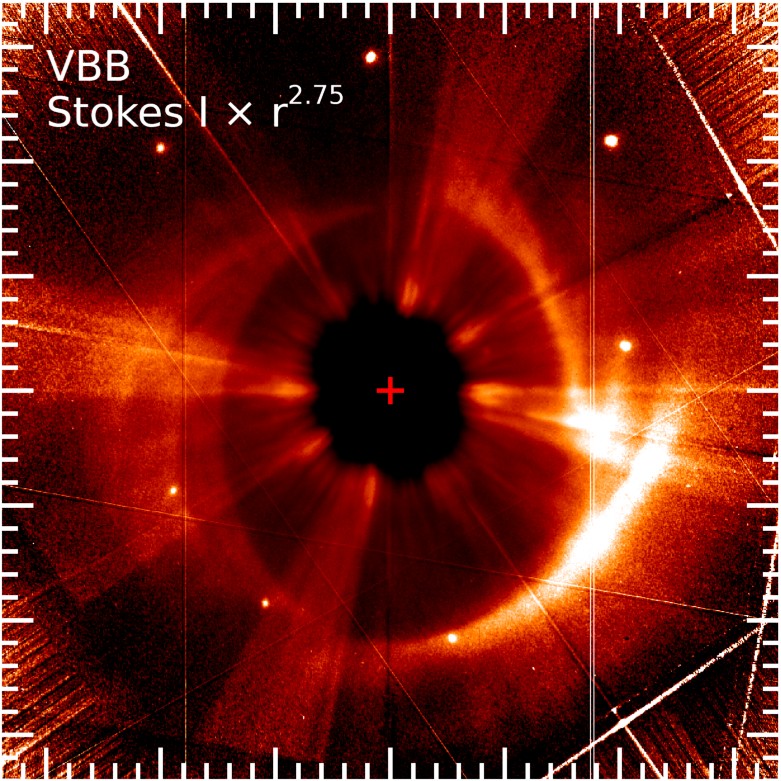} \\
\includegraphics[totalheight=3.3in]{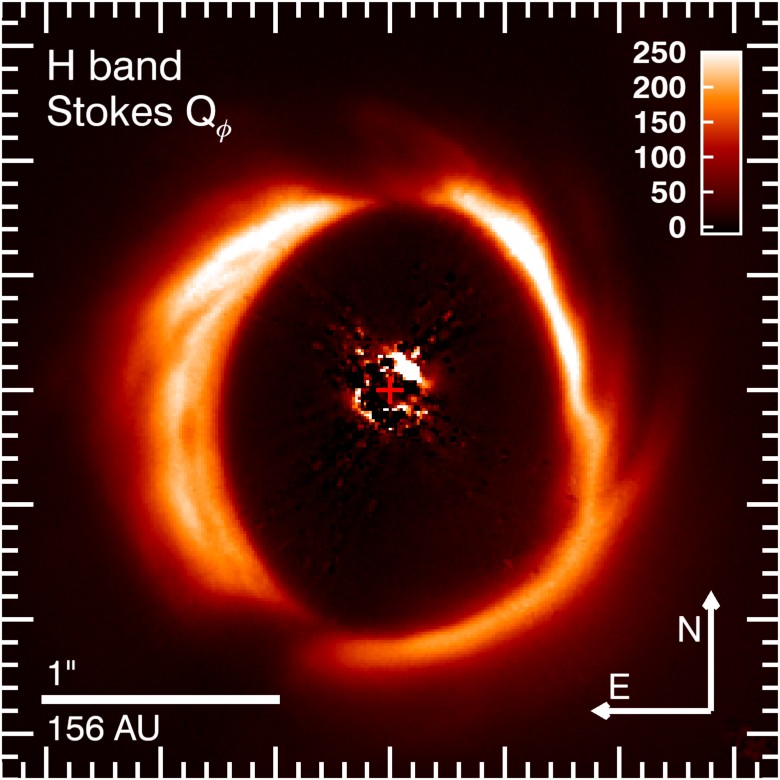} & \includegraphics[totalheight=3.3in]{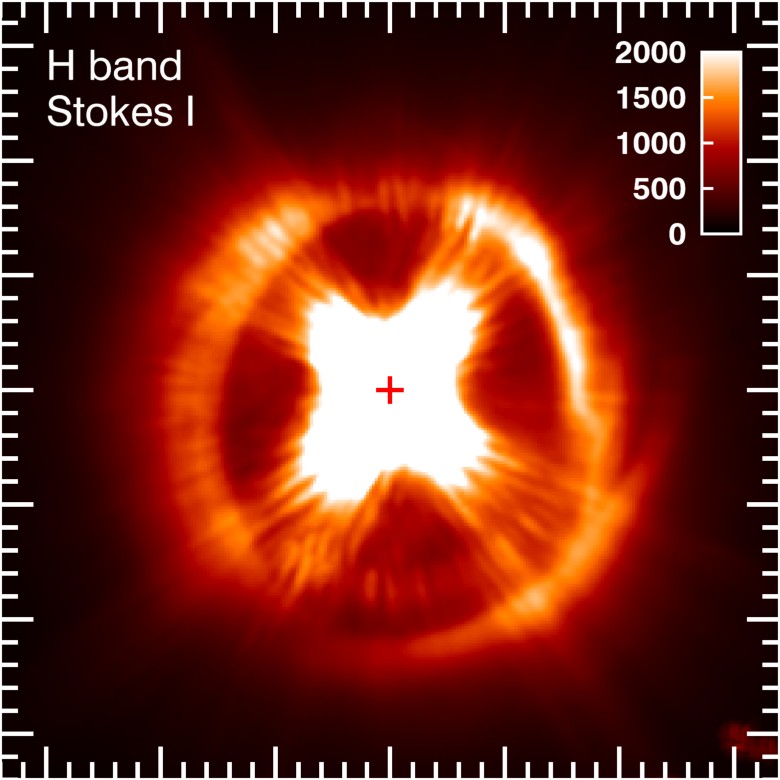} \\
\end{tabular}
\caption{Polarized intensity $Q_{\phi}$ and total intensity $I$ signal of the disk around HD142527 for the non-coronagraphic VBB and the coronagraphic $H$-band observations. The scale for the size and orientation for all images is shown in the bottom left image. The position of the star is marked with a cross. The color scales are in counts per DIT. The VBB Stokes $I$ image in the top right frame is scaled with $r^{2.75}$ ($r$ is the distance from the star) to improve the visibility of the disk, and therefore no color scale is given. The bright point-like spots in the same image are ghost images of the star that are produced by the instrument.}
  \label{fig:HD142527_stokes_I_Qphi}
\end{figure*}

The SPHERE/IRDIS observations were obtained on 2017 May 31 in field-stabilized dual-beam polarimetric imaging (DPI) mode using a classical Lyot coronagraph and the broadband $H$ filter (BB\_H) with central wavelength $\lambda_c=1625.5~{\rm nm}$ and bandwidth $\Delta\lambda=291~{\rm nm}$. The DITs during the observations were alternated between 8~sec and 16~sec. For most measurements in this work, we used the combined result with both DITs. The flux calibration observations were made with short DIT~=~2~sec exposures with the neutral density filter ND\_2.0 that diminishes the flux by about a factor of 100.

Because these observations were taken during the same night as the third ZIMPOL epoch, they also show the low-wind effect, which affects the resolution of the data. However, the effect on the $H$-band dataset is not very disturbing because the limited resolution has no significant effect on the signal extraction of an extended bright circumstellar disk. In addition, the brightness of the disk with respect to the star is much higher in the $H$-band than in the VBB.

The reference dataset for the shape of the IRDIS $H$-band PSF without disk is the polarimetric imaging of HR8799 from 2016 October 11. This dataset was published in \citet{vanHolstein17}, and we used it here because it does not contain any strong signals\footnote{The planets orbiting HR8799 are too faint compared to the bright disk in HD142527 to interfere with the analysis, and their signals are additionally suppressed by the RDI process.} other than the stellar PSF and was obtained during similar conditions as the science data. In addition, both datasets are coronagraphic, and the stars have a similar brightness in $H$-band. The post-processed HR8799 dataset contains a total of 172 different $H$-band PSFs, which we all used as reference frames in our calculations. More detailed information about the IRDIS datasets can be found in Table~\ref{table: data}.

The data reduction for the IRDIS data was performed with the end-to-end IRDAP data reduction pipeline \citep{vanHolstein20}. All frames were centered on a common estimated mean center, which was determined with the four satellite spots of the star center frames \citep{Beuzit19}. In addition to the standard calibration steps and centering of the frames, IRDAP also corrects the polarimetric crosstalk, removes instrument polarization, and measures and removes interstellar polarization and the intrinsic polarization of the star (see discussion in Sec.~\ref{sec:Intrinsic stellar polarization}).

The image format of the IRDIS data after the reduction is $1024\times1024$ pixels, with a pixel size of $12.27\times12.27$~mas. We determined a resolution of 44~mas (see Fig.~\ref{fig:data_prof_reso}) for these data. This value is slightly lower than expected for the observing conditions listed in Table~\ref{table: data} because of the low-wind effect. The final reduced images, intensity (Stokes $I$), and polarized intensity (Stokes $Q_{\phi}$) for the $H$-band observations are shown in Fig.~\ref{fig:HD142527_stokes_I_Qphi}.

\subsection{Intrinsic stellar polarization}
\label{sec:Intrinsic stellar polarization}
As part of the data reduction for all datasets, we normalized the $Q$ and $U$ polarization measured in an aperture for each individual exposure by adjusting the scale factors $c_Q$ and $c_U$ in the subtraction $Q_{\rm norm} = Q-c_Q\cdot I = 0$ and similar for Stokes $U$. We used an annulus from $r=0.2\arcsec-0.6\arcsec$ as aperture, containing much light from the central source, but avoiding the saturated center and the scattered light from the inner and outer disks. This normalization procedure is common practice for differential polarimetric imaging of disks to remove disturbing polarization signals from the telescope and the interstellar polarization because a disk often only becomes visible after this polarimetric normalization \citep[e.g.,][]{Quanz11, Avenhaus14}. However, the method assumes that the central source has no intrinsic polarization, and this is questionable for HD142527 because there is polarization at a level of 0.5-1.0\% from a very compact, partly unresolved inner disk \citep{Avenhaus17}, and we also obtained unusually large $c_Q$ and $c_U$ normalization factors. 

\subsubsection{Correction for the telescope polarization}
The telescope polarization is mainly introduced by inclined mirrors and strongly depends on the telescope pointing direction. This has been modeled for VLT/SPHERE for measurements of polarization standard stars with a precision of $\Delta p \approx \pm 0.1 \%$ \citep{Schmid18,vanHolstein20}. For the ZIMPOL data, a precise measurement is possible because the observations of 2017 exhibit a large parallactic angle range of about $50^{\circ}$ or a circular arc of $100^{\circ}$ in the Q/I-U/I plane, as discussed in detail in Appendix A.2 of \citet{Hunziker20}.

We measured a total sum of intrinsic and interstellar polarization of $p_{\star}({\rm VBB})=1.0\pm0.1 \%$ for the 2017 data along a position angle $\varphi_{\star}({\rm VBB})=51\pm3 ^{\circ}$ and $p_{\star}(H)=0.6\pm0.1 \%$, $\varphi_{\star}(H)$ of $69\pm5 ^{\circ}$. The 2015 and 2016 VBB data are not optimal for measuring the stellar polarization because the observations were taken after meridian passage with small changes in parallactic angles. In addition, the telescope mirrors were recoated later in 2016, and therefore the telescope polarization is much higher for the 2015 and 2016 data than for the 2017 data. However, by combining all the 2015 and 2016 data, we were able to determine $p_{\star}({\rm VBB})=0.8\pm0.1 \%$ and $\varphi_{\star}({\rm VBB})=49\pm4 ^{\circ}$, which agrees well with the measurements from 2017. To our knowledge, there are no prior measurements published for the linear polarization of HD142527.

This total stellar polarization could be caused by interstellar polarization from absorption by magnetically aligned dust grains along the line of sight or by intrinsic stellar polarization and scattering from hot dust near the star, which was partly resolved by \citet{Avenhaus17}. However, the polarization in this paper was also normalized, and the net contribution of the star in the inner disk ($<0.2\arcsec$) to the polarization in the annulus that was used for the normalization is not known.

The interstellar polarization contributes a factor $c_{\rm ism}$ to the stellar polarization, which applies equally for the star and the disk. If the total stellar polarization is only caused by interstellar polarization, then the normalization would correct for it, as for the telescope polarization. However, if there is a significant contribution from the central object to the polarization, then the normalization is nulling this real signal from the central object, and an error in the normalized polarization signal of the disk is introduced. Therefore we investigate the nature of the total stellar polarization for HD142527.

\subsubsection{Interstellar polarization for HD142527}
A polarization of $p_{\star}({\rm VBB})\approx 1.0~\%$ and $p_{\star}(H)\approx 0.55~\%$ would be a high interstellar polarization for a star at a distance of $156~{\rm pc}$ like HD142527. Other stars at similar distances only exhibit polarizations of a few dozen percent, and polarizations of $>0.5 \%$ are uncommon for distances <$500$~pc \citep{Gontcharov19}. However, our object is located in the Lupus cloud complex, a star-forming region at a distance of $\sim$150--200~pc \citep{Comeron08, GAIA18}. Its exact location is close to the Lupus 4 field, one of the large dusty clouds in Lupus, and other stars in this field show high linear polarizations $p\approx 0.5-2.5\%$ in the optical with a rather narrow distribution of position angles of $\varphi_p=26\pm 9^{\circ}$ \citep{Rizzo98}, which can be explained by a large-scale interstellar polarization component.

\citet{Pottasch88} estimated the extinction from dust along the line of sight toward HD142527 to be $A_V = 0.6$, which was supported by \citet{Verhoeff11}, who also obtained an extinction of $A_V = 0.6$  to account for the reddening of the stellar light in their model. However, \citet{Malfait98} noted a discrepancy between the observed total extinction and the interstellar extinction in
their photometric study, indicating that potentially variable intrinsic dust contributes along the line of sight.

In addition, the position angles $\varphi_{\star}$ for the polarization $p_{\star}$ of HD142527 in our data are about $25^{\circ}$ larger for the VBB and about $43^{\circ}$ larger in the $H$-band than the mean presented in \citet{Rizzo98}, indicating that an intrinsic polarization component contributes as well, particularly for the $H$-band. Moreover, the measured wavelength dependence of the total stellar polarization $\Lambda=p_{\star}(H)/p_{\star}(VBB)=0.6\pm 0.1$ does not agree well with interstellar polarization as described by \citet{Serkowski75}, which predicts a lower ratio of $\Lambda \approx 0.3$\footnote{Assuming $p_{\rm max}$ to be located at wavelength $\lambda_{\rm max} = 545~{\rm nm}$ and $K=1.15$} on average, again indicating that an intrinsic polarization component is present in HD142527. However, this is based on an empirical law with significant uncertainties and should therefore not seen as strong evidence.

The combined evidence shown in this section points towards the presence of interstellar and intrinsic dust components which can both contribute to the observed polarization $p_{\star}$ of HD142527, but with the limited available data from previous studies it is not possible to disentangle the two polarization components. Therefore, we have to assume that the star exhibits at least some intrinsic polarization in addition to the interstellar polarization. Spectropolarimetry of this object would be required to determine the contributions of intrinsic and interstellar polarization. 

\subsubsection{Polarimetric normalization and disk polarization}
\begin{figure}
\centering
\includegraphics[width=\linewidth]{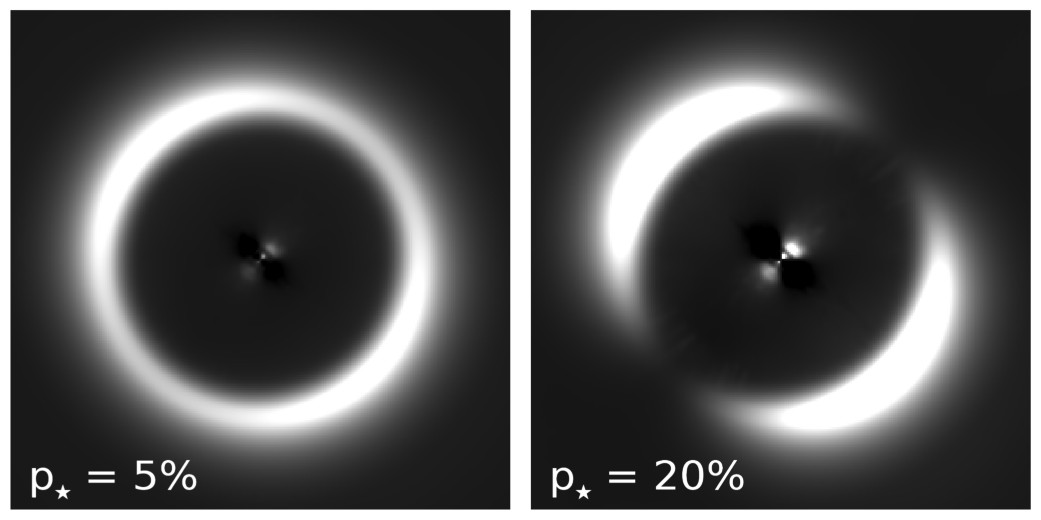} \\
\caption{Simulated Q$_{\phi}$ disk signal for an azimuthally symmetric disk with $p_{\rm disk}=20\%,$ illustrating the effect of polarimetric normalization for a star with very strong (10 to 40 times stronger than estimated for HD142527) intrinsic polarization of 5\% (left) and 20\% (right) with $\varphi_{\star} = 45^{\circ}$.}
\label{fig:HD142527_200pos_disk_sim_Hband_inspol}
\end{figure}

We investigated the effect of the applied normalization on the disk signal for HD142527 because we have strong evidence that the star and the barely resolved inner hot dust produce a net intrinsic polarization of $p\approx 0.3-0.7~\%$. A detailed calculation of how the normalization of $Q$ and $U$ affects polarized signals close to the star for an intrinsically polarized star is presented in \citet{Hunziker20} (Appendix~B). This formalism can also be applied to the case of an extended source, in particular, the disk of HD142527. We assumed a disk with intrinsic polarized signals $Q^{\rm int}_{\phi, \rm disk} = p_{\rm disk} I_{\rm disk}$ and $U^{\rm int}_{\phi, \rm disk} = 0$ and a star with intrinsic polarization components $p_{\star, Q}$ and $p_{\star, U}$, and applied Eq.~(B.8) from \citet{Hunziker20} to determine the disk signal $Q_{\phi, \rm disk}$ and $U_{\phi, \rm disk}$ after the normalization,

\begin{equation}
\begin{aligned}
Q_{\phi, \rm disk}/I_{\rm disk} &= p_{\rm disk} + p_{\star, Q}\cos\left(2\phi\right) + p_{\star, U}\sin\left(2\phi\right)\\
U_{\phi, \rm disk}/I_{\rm disk} &= - p_{\star, Q}\sin\left(2\phi\right) + p_{\star, U}\cos\left(2\phi\right).
\end{aligned}
\label{equ: normalized disk signal}
\end{equation}

This shows that the measured degree of polarization for the disk $p_{\rm disk}$ changes due to the normalization by an amount corresponding to the intrinsic stellar polarization $p_{\star}$. This means that even a large uncertainty in $p_{\star}$ of $0.1\%$ for the stellar polarization would only have a small effect on the relative disk polarization $Q_{\phi, \rm disk}/I_{\rm disk}$ if $p_{\rm disk}$ is at a level of $20-40\%,$ as for HD142527. Furthermore, the change is not constant over the whole image, but for $Q_{\phi, \rm disk}$ it adds a quadrant pattern aligned with the direction $\varphi_{\star}$. A quadrant pattern rotated by $45^\circ$ is also added to the $U_{\phi, \rm disk}$ signal, so that $U_{\phi, \rm disk}$ systematically deviates from zero. A nonzero $U_{\phi, \rm disk}$ signal would usually not be expected for a large low-inclination disk like HD142527 unless the scattering grains are aligned or second-order scattering effects are really strong, which was investigated by \citet{Canovas14} for an unpolarized central source. The amplitude of the introduced pattern is $p_{\star}I_{\rm disk}$ for both $Q_{\phi, \rm disk}$ and $U_{\phi, \rm disk}$.

To illustrate the effect, we simulated an azimuthally symmetric disk with $p_{\rm disk}=20\%$ and a star that exhibits intrinsic polarization along the direction of $U$ (or $\varphi_{\star} = 45^{\circ}$), which is similar to HD142527, but with an unreasonably high $p_{\star}$ of $5\%$ and $20\%$ for illustration purposes. In Fig.~\ref{fig:HD142527_200pos_disk_sim_Hband_inspol} we show $Q_{\phi, \rm disk}$ after normalizing $Q$ and $U$ in a circular aperture $<0.6\arcsec$ that did not overlap with the disk signal. The $Q_{\phi, \rm disk}$ signal after the normalization clearly deviates from azimuthal symmetry, with the deviation aligned with the direction $\varphi_{\star}$ and a variation of $p_{\rm disk}$ of about $\pm p_{\star}$, in agreement with Eq.~(\ref{equ: normalized disk signal}).

The polarization of the disk of HD142527 is affected in a similar way as the disk model shown in Fig.~\ref{fig:HD142527_200pos_disk_sim_Hband_inspol}. Our measured polarization $p_{\rm disk}$ along the position angle of $p_{\star}$ (around $60^{\circ}$) should be artificially increased by $\lesssim1\%$ and decreased by the same amount in the perpendicular direction (around $150^{\circ}$) if $p_{\star}$ is purely due to intrinsic stellar polarization. Because our resolved measurements yield typical values of $p_{\rm disk} \approx 30\%$ with errors larger than $1\%$, we consider the effect of the possible intrinsic polarization of the star as negligible. Other more accurate measurements such as the total disk-integrated degree of polarization $p_{\rm disk}$ are less affected because the positive and negative overcorrections introduced by the normalization mostly average out when integrated over the whole or over half of the disk because of the azimuthal periodicity of the induced variation. However, because the disk is slightly asymmetric, we find that the normalization can result in a maximum error of about $0.3 p_{\star}$ for the total disk-integrated degree of polarization $p_{\rm disk}$. This corresponds to an error of $0.3\%$ and $0.17\%$ for the VBB and $H$-band data, respectively.

\section{Data analysis}
\label{sec:Data analysis}
Our measurements include the integrated total intensity of the system $I_{\rm total}$ in the VBB and $H$-band and the total intensity $I_{\rm disk}$ and polarized intensity $Q_{\phi, \rm disk}$ from the disk alone. In addition, we determined the intensity $I_{\rm disk}^{\rm far}$ and polarized intensity $Q_{\phi, \rm disk}^{\rm far}$ from the far side of the disk as a simple way to quantify the brightness difference between the predominantly forward- and backward-scattering side. Furthermore, we measured radial profiles of the disk intensity $I_{\rm disk}(\varphi, r)$ and polarized intensity $Q_{\phi, \rm disk}(\varphi, r)$, with $\varphi$ the position angle (PA) with respect to north and $r$ the projected separation from the star. We then used the measurements of $Q_{\phi}$ and $I$ to derive the linear degree of polarization $p = Q_{\phi}/I$ of the reflected light. This section describes the details of our data analysis. The most relevant measurement results are then transformed into physical units and are summarized in Section~\ref{sec:precise measurements}.

\subsection{PSF smearing effect}
\label{sec:PSF smearing effect}
\begin{figure}
\centering
\includegraphics[width=\linewidth]{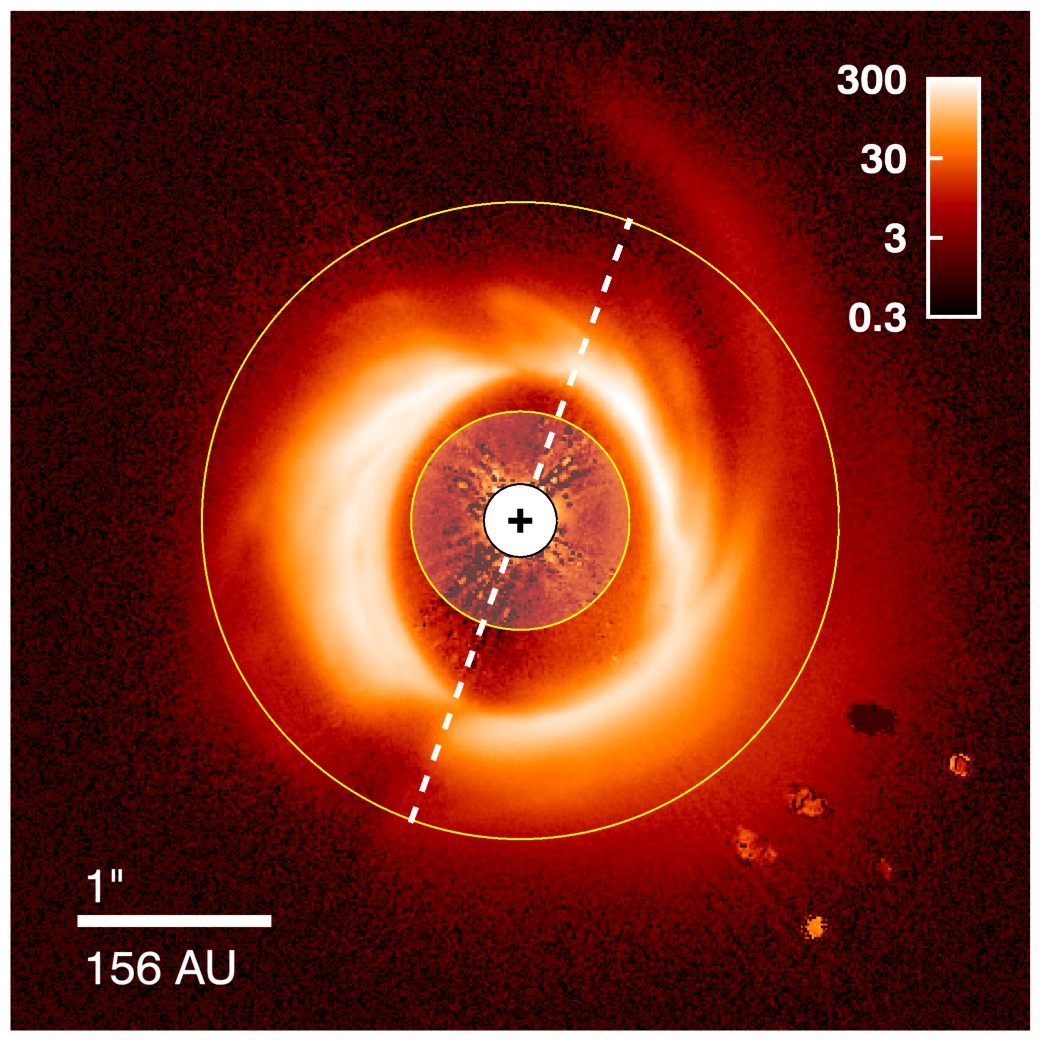}
\caption{HD142527 $Q_{\phi}$ image of the outer disk in $H$-band with IRDIS. The two yellow circles at 0.6\arcsec and 1.75\arcsec \ indicate the annulus aperture we used to measure the total disk intensity. A white disk with radius 0.2\arcsec covers the inner part of the image, which is dominated by residuals from speckles and from the coronagraph. The position of the star behind the disk is marked with a cross. The dashed line indicates the approximate direction of the disk major axis at a position angle of -20$^{\circ}$. The scale is in counts for a 16~sec DIT. North is up and east is to the left. The features to the southwest of the disk originate from defects on the IRDIS detector.}
  \label{fig:HD142527_stokes_aper_Qphi_IRDIS}
\end{figure}

The spatial resolution of the observations is determined by the wavelength and by the performance of the AO system and the observing conditions. The finite resolution has a strong effect on measurements in small apertures. We refer to this as the PSF smearing effect. An analysis of the unsaturated PSF for the VBB observations shows, for example, that only about 50\% of the total intensity is located within $3\lambda/D$ of the center of the PSF of a point source \citep[see, e.g.,][]{Schmid18}. Therefore a significant fraction of the total intensity is omitted in photometry with smaller apertures unless the measured photometric flux values are corrected. In addition, the effect is different for the $I$ and $Q_{\phi}$ signal. The $I$ signal is just smeared because it is strictly positive. For the $Q$ and $U$ signal, however, the PSF smearing additionally leads to crosstalk between disk regions with negative and positive polarization, which cancels out a fraction of $Q_{\phi}$ \citep{Schmid06b}. This additional effect must be considered to avoid incorrect results for the degree of polarization $Q_{\phi}/I$. In order to estimate correction factors $f_{\rm corr}$ for our measurements, we simulated these effects in the following way: We first simulated $I$, $Q_{\phi}$, and $U_{\phi}$ images of a face-on azimuthally symmetric disk with $U_{\phi}=0$ and with radial profiles similar to the real disk. We then transformed the $Q_{\phi}$ and $U_{\phi}$ images into the $Q$ and $U$ basis and convolved all intensities $I \rightarrow \tilde{I}$, $Q \rightarrow \tilde{Q}$ and $U \rightarrow \tilde{U}$ with the unsaturated PSF of the particular observations\footnote{We use $\text{the tilde}$ to indicate that a value is affected by PSF smearing. Whenever the symbol is not used for a measurement, the value is already corrected for PSF smearing.}. Afterwards we transformed the convolved $\tilde{Q}$ and $\tilde{U}$ images back into the $\tilde{Q}_{\phi}$, $\tilde{U}_{\phi}$ basis and finally applied all measuring procedures to $I$, $Q_{\phi}$ and $\tilde{I}$, $\tilde{Q}_{\phi}$, and calculated the correction factors $f_{\rm corr}^I = I/\tilde{I}$ and $f_{\rm corr}^P = Q_{\phi}/\tilde{Q}_{\phi}$ for the photometric results.

We measured the uncorrected integrated intensities of the disk $\tilde{I}_{\rm disk}$ and $\tilde{Q}_{\phi, \rm disk}$ in an annular aperture from 0.6\arcsec to 1.75\arcsec around the central star as indicated in Fig.~\ref{fig:HD142527_stokes_aper_Qphi_IRDIS}. The outer radius was chosen as large as possible to fit into the field of view (FOV) of ZIMPOL. We investigated the amount of integrated intensity lost due to PSF smearing for simulated disks with radial profiles similar to the disk in HD142527, and we report the correction factors $f_{\rm corr}$ for the integrated disk intensities resulting from this analysis in Table~\ref{table: PSF smearing correction factors}. The values in Table~\ref{table: PSF smearing correction factors} show the average correction factors for all of the tested radial profiles, and the errors show the variance for different radial profiles. We find that $f_{\rm corr}^I$ and $f_{\rm corr}^P$ are different, and there is no strong dependence on the radial profile of the disk. However, we note that more integrated intensity is lost in general (higher $f_{\rm corr}$ values) for disk profiles that are less strongly radially extended. The effect becomes increasingly stronger for small disks that are not much larger than the data resolution. The values in the table show that $f_{\rm corr}$ is significantly higher for the VBB data than for the $H$-band data, even though the resolution in VBB is better. This shows that the effect of the Strehl ratio, which was lower in the visible, is stronger in this case because the PSFs in observations with low Strehl ratio have a strong residual halo (Fig.~\ref{fig:data_prof_reso}), therefore the total flux of the PSF is smeared over a larger portion of the image.

\begin{table}[h]
\caption{Correction factors for PSF smearing for the HD142527 observations}
\label{table: PSF smearing correction factors}
\centering
\begin{tabular}{llll}
\hline\hline
Measurement & Filter & $f_{\rm corr}^I$ & $f_{\rm corr}^P$ \\
\hline
\multirow{2}{*}{annulus} & VBB & $1.12\pm0.01$ & $1.30\pm0.02$ \\
 & $H$-band & $1.07\pm0.01$ & $1.18\pm0.01$\\
\hline
\multirow{2}{*}{radial profiles} & VBB & $1.67\pm0.14$ & $1.67\pm0.14$ \\
 & $H$-band & $1.35\pm0.07$ & $1.35\pm0.07$\\
 \hline
\end{tabular}
\end{table}

\begin{figure}
\centering
\includegraphics[width=\linewidth]{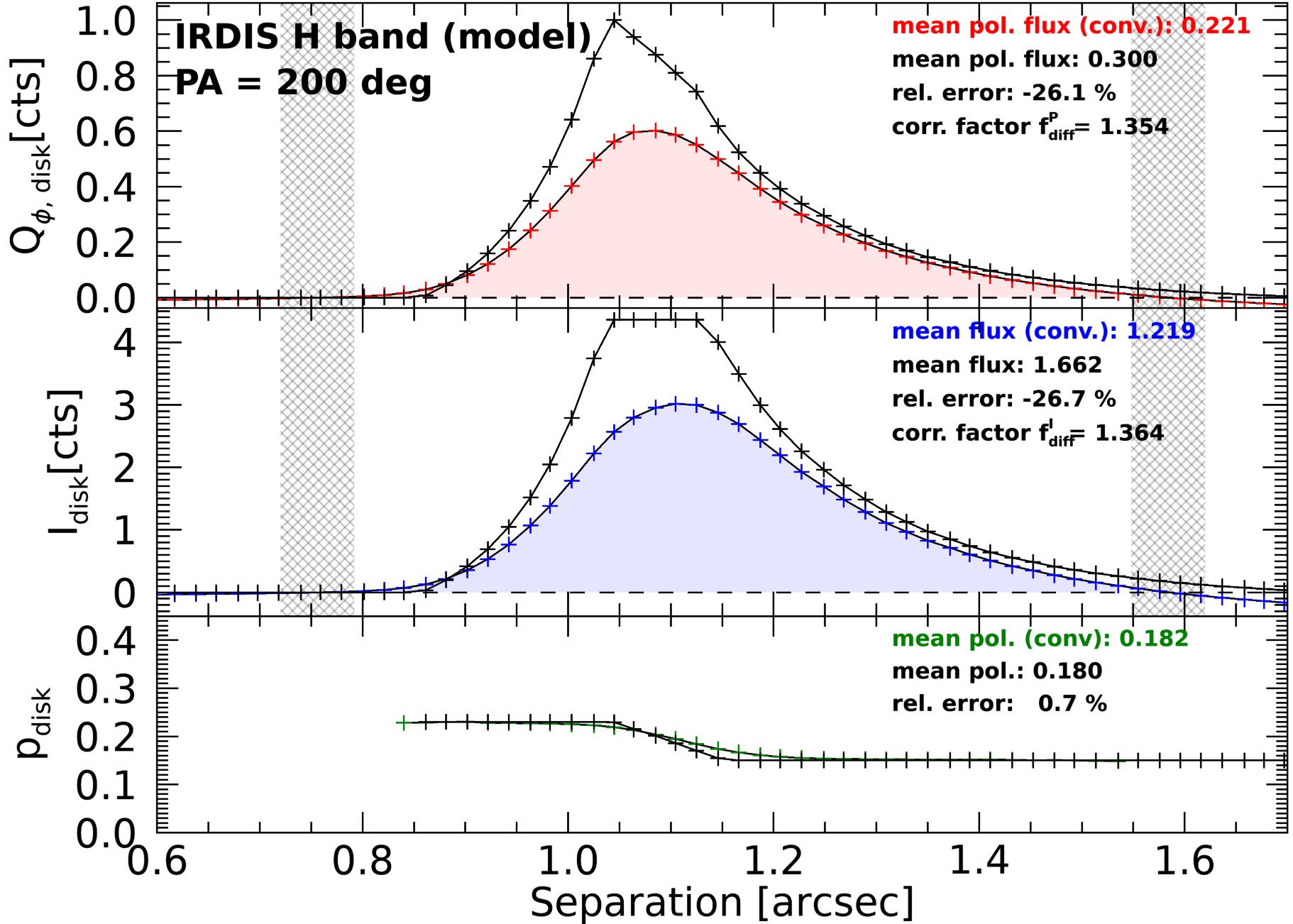} \\
\caption{Simulated radial profiles for Q$_{\phi}$ (red) and Stokes I (blue) and the degree of polarization Q$_{\phi}/$I (green) convolved with the IRDIS $H$-band PSF. The black curves show the initial model profiles of the disk before the convolution. The convolved profiles roughly correspond to the disk measurements at position angle 200$^{\circ}$ (see Fig.~\ref{fig:HD142527_200pos_profiles}, bottom). The radially integrated mean disk intensities and the relative lost intensity (rel. error) were calculated for both profiles between the cross-hatched areas.}
\label{fig:HD142527_200pos_sim_profiles_Hband}
\end{figure}

In the same way, we modeled the effect of PSF smearing for the measurements of the radial disk intensity profiles $\tilde{I}_{\rm disk}(\varphi, r)$ and $\tilde{Q}_{\phi, \rm disk}(\varphi, r)$ and report the $f_{\rm corr}$ values in Table~\ref{table: PSF smearing correction factors}. Because the profile measurements are different from the total intensity integration, the resulting $f_{\rm corr}$ are different as well. The corrections are significantly larger because a larger fraction of the "smeared" flux is missed in a radial line extraction compared to the integration in a large aperture. This is further discussed in Sec.~\ref{sec:Radial profile analysis}. The advantages of the radial profiles are the equal correction factors $f_{\rm corr}^I = f_{\rm corr}^P$, which result in much more precise measurements of the degree of polarization because $p = Q_{\phi}/I = \tilde{Q}_{\phi}/\tilde{I}$ eliminates one major source of error. The situation is depicted in Fig.~\ref{fig:HD142527_200pos_sim_profiles_Hband} for simulated $H$-band disk intensity profiles similar to the measured profiles at $\varphi = 200^{\circ}$, a position angle where the profile determinations are very accurate for both wavelengths. We show the convolved profiles $\tilde{Q}_{\phi, \rm disk}(\varphi, r)$, $\tilde{I}_{\rm disk}(\varphi, r)$ and $\tilde{p}_{\rm disk}(\varphi, r) = \tilde{Q}_{\phi, \rm disk}(\varphi, r)/\tilde{I}_{\rm disk}(\varphi, r)$ together with the initial model profiles before convolution. The radially integrated intensity profiles $Q_{\phi}$ and $I$ of the convolved disk signal underestimate the true model intensities by about 26\% on average, which corresponds to  $f_{\rm corr}^I = f_{\rm corr}^P \approx 1.35$. However, the measured degrees of polarization of the convolved disk $\tilde{p}_{\rm disk}(\varphi, r)$ are very close to the initial model values at all separations $r$ because the measured profiles $\tilde{I}_{\rm disk}(\varphi, r)$ and $\tilde{Q}_{\phi, \rm disk}(\varphi, r)$ are both degraded in the same way by the PSF smearing (see Sec.~\ref{sec:Radial profile analysis}).

\subsection{Integrated intensity analysis}
\label{sec:Integrated disk flux analysis}

First, we integrated the polarized disk intensity $Q_{\phi, \rm disk}$ in an annular aperture from $0.6\arcsec$ to $1.75\arcsec$ relative to the total intensity of the system $I_{\rm total}$ in a circular aperture with radius 1.75\arcsec (see Fig.~\ref{fig:HD142527_stokes_aper_Qphi_IRDIS}). We used the flux calibration frames of the $H$-band observations to measure $I_{\rm total}({H})$ and the unsaturated VBB observations from 2016 to measure $I_{\rm total}({\rm VBB})$. We obtained $I_{\rm total}({\rm VBB})=4.13\pm0.02 \cdot 10^6~{\rm cts}/{\rm sec}$ and $I_{\rm total}({H})=2.00\pm0.10 \cdot 10^7~{\rm cts}/{\rm sec}$ for the total system intensities and $Q_{\phi, \rm disk}/I_{\rm total} ({\rm VBB}) = 0.46\pm0.02\%$ and $Q_{\phi, \rm disk}/I_{\rm total} (H) = 1.07\pm0.07\%$ for the polarized disk intensity contrast. We assume that the $Q_{\phi, \rm disk}/I_{\rm total}$ values are constant in time because no sign of variability has been reported up to now for this disk. The errors are dominated by the uncertainty of the correction factors in Table~\ref{table: PSF smearing correction factors}.

Measuring the disk intensity is difficult because we have to separate the weak disk signal $I_{\rm disk}$ from the stellar PSF in the Stokes $I$ image. For a rough estimate of $I_{\rm disk}$, we scaled the reference VBB and $H$-band PSFs to the brightness of HD142527 and removed it from the Stokes $I$ image of the respective HD142527 dataset. This analysis resulted in intensity contrast values of $I_{\rm disk}/I_{\rm total}({\rm VBB})<4$\% and $I_{\rm disk}/I_{\rm total}(H)=4\pm1$\%. The errors are dominated by the variability of the reference PSFs. For the VBB data, it was not possible to extract the disk intensity accurately, resulting in a mere upper limit for the contrast. More accurate results for the total disk intensity measurements follow from the degree of polarization in Sec.~\ref{sec:precise measurements}.

We also quantified the polarized intensity for the near and far sides of the disk inside the annular aperture. For position angles $\varphi$ ranging from $-20^{\circ}$ to $160^{\circ}$, that is, the eastern side of the annulus shown in Fig.~\ref{fig:HD142527_stokes_aper_Qphi_IRDIS}, we measure $Q_{\phi, \rm disk}^{\rm far}/I_{\rm total} ({\rm VBB}) = 0.23\pm0.01\%$ in the
visual, or in other words, about $49\pm4\%$ of the polarized disk intensity originates from the far side of the disk. In the $H$-band, the polarized intensity from the far side is $Q_{\phi, \rm disk}^{\rm far}/I_{\rm total} (H) = 0.56\pm0.05\%,$ which corresponds to $52\pm9\%$ of the total polarized intensity from the disk. The results suggest that the reflected polarized intensity is evenly distributed on the near and far sides of the disk for both wavelengths. These values were already corrected for PSF smearing using the values from Table~\ref{table: PSF smearing correction factors}.

\subsection{Radial profile analysis}
\label{sec:Radial profile analysis}
\subsubsection{Disk polarized intensity profiles}

\begin{figure}
\centering
\includegraphics[width=\linewidth]{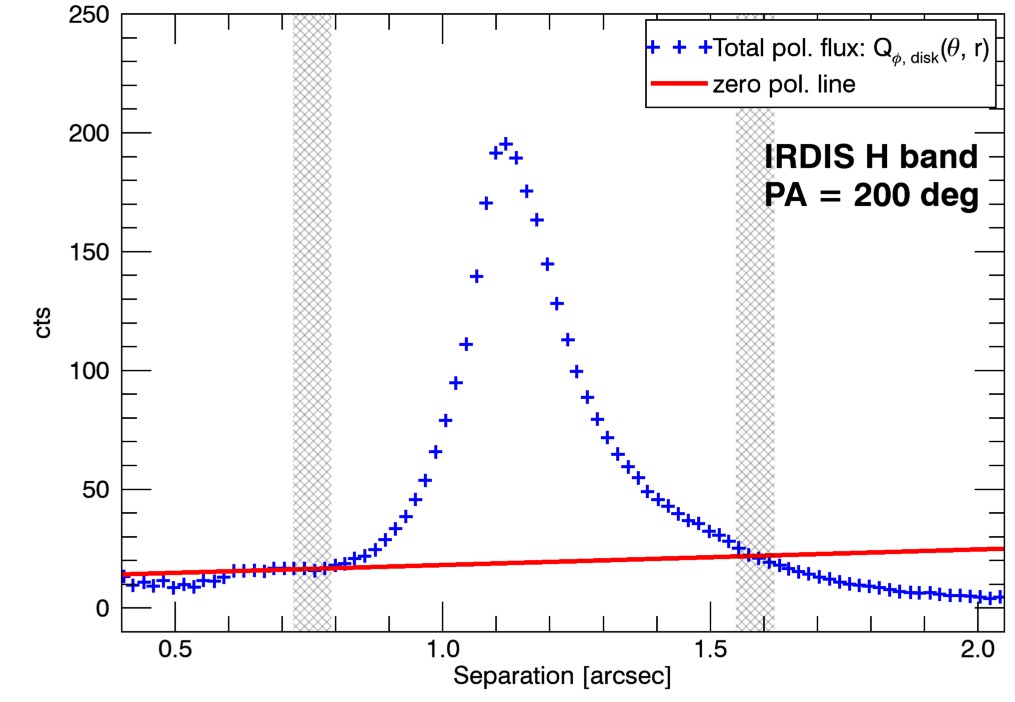}
\caption{Measured radial polarized intensity profile from the HD142527 IRDIS data in $H$-band (blue crosses) at a position angle of $200^{\circ}$. The solid red line was fit through the data points in the cross-hatched areas. We consider this as the zero line for the polarized surface brightness radial profiles.}
  \label{fig:HD142527_IRDIS_PolBG_200pos_20avg_20pix}
\end{figure}

\begin{figure}
\centering
\begin{tabular}{cc}
\includegraphics[width=\linewidth]{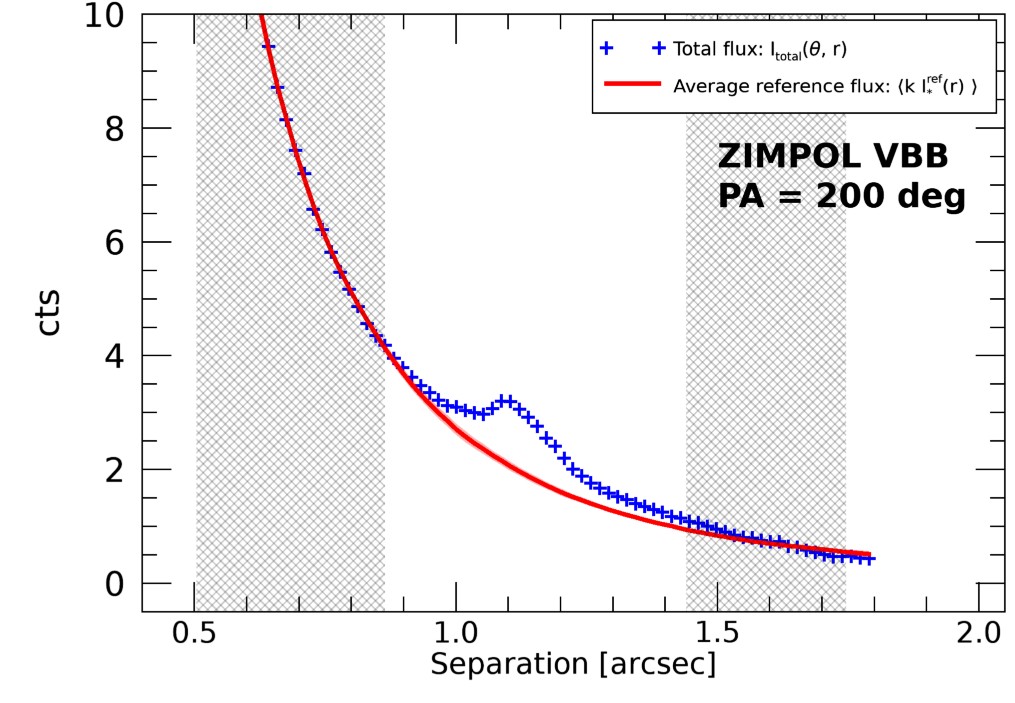} \\
\includegraphics[width=\linewidth]{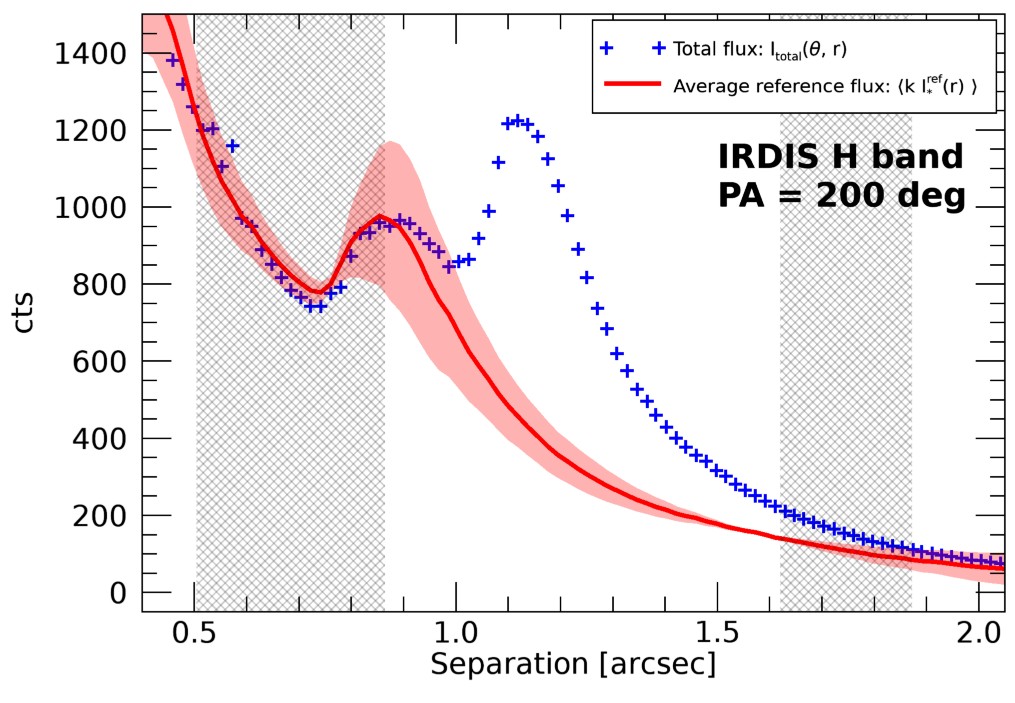} \\
\end{tabular}
\caption{Measured radial intensity profiles of HD142527 for the ZIMPOL VBB and the IRDIS $H$-band data (blue crosses) at a position angle of $200^{\circ}$ compared to the mean of the best-fitting radial profiles from the reference data (solid red line). The shaded red area shows the 1$\sigma$ spread of the reference profiles. All reference profiles were fit only in the cross-hatched areas to the HD142527 data to avoid fitting and subtracting the disk signal.}
  \label{fig:HD142527_PSFbestfits_200pos}
\end{figure}

\begin{figure}
\centering
\begin{tabular}{l}
\includegraphics[width=\linewidth]{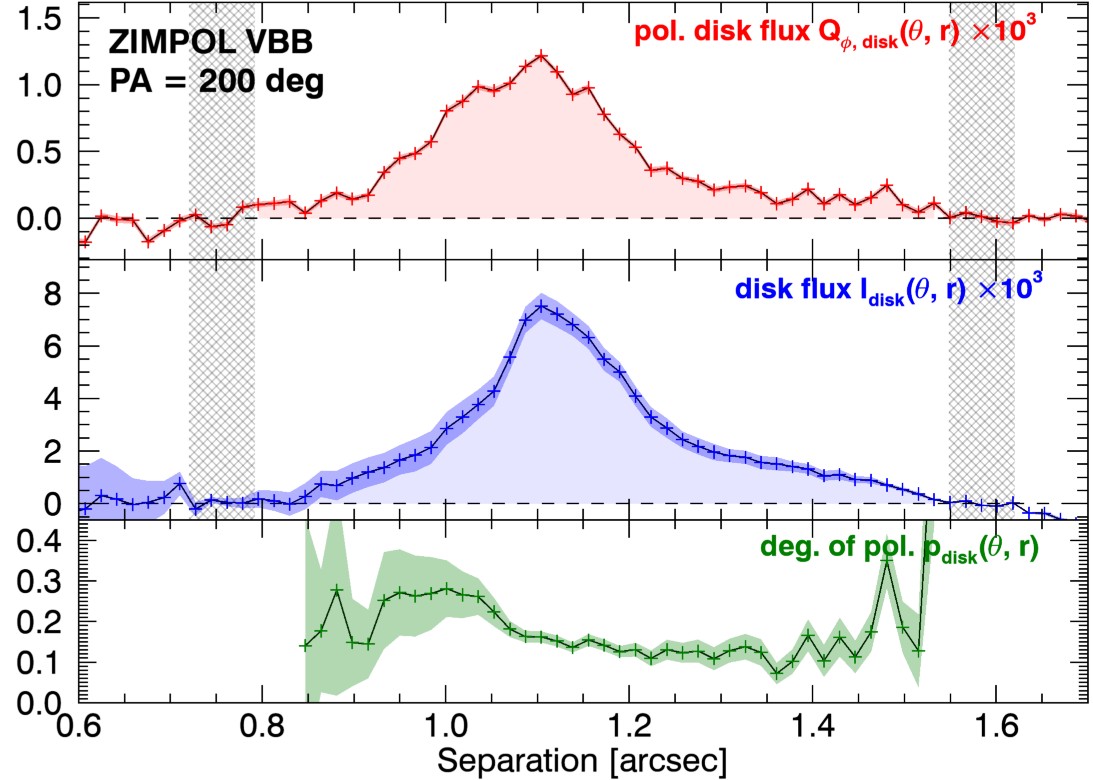} \\
\includegraphics[width=\linewidth]{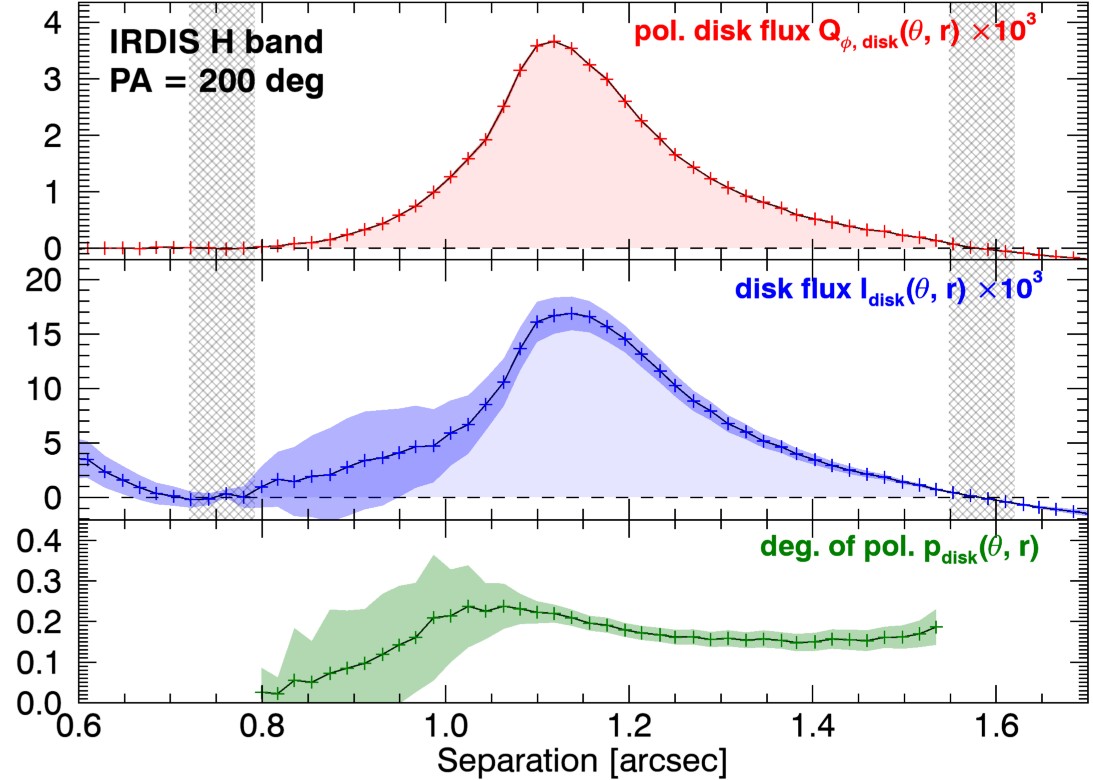} \\
\end{tabular}
\caption{Radial profiles for Q$_{\phi}$ (red), Stokes I (blue), and the degree of polarization Q$_{\phi}/$I (green) at a position angle of $200^{\circ}$ for the ZIMPOL VBB observations (top) and for the IRDIS $H$-band observations (bottom). The intensity values on the vertical axis are in units of surface brightness relative to the total intensity of the system, i.e., [$I_{\rm total}/{\rm arcsec}^2$]. For the degree of polarization, only the values inside the cross-hatched boundaries with <0.5 error are shown in the plot to improve the visibility.}
  \label{fig:HD142527_200pos_profiles}
\end{figure}

For the measurement for each radial profile $\tilde{Q}_{\phi, \rm disk}(\varphi, r)$, we first extracted a line of pixels in the $Q_{\phi}$ images along a position angle $\varphi$, starting from the position of the star at $r=0$. The pixels were then binned to lower the statistical errors of the measurement, but we retained a sampling in radial direction of $\sim0.015\arcsec$. For each profile we defined the positions of the inner and outer edge of the disk signal and determined the mean intensity of several pixels at these points. We used a straight line through these points as zero level for $\tilde{Q}_{\phi, \rm disk}(\varphi, r)$, just as indicated in Fig.~\ref{fig:HD142527_IRDIS_PolBG_200pos_20avg_20pix}. This removes the polarized intensity outside the two cross-hatched areas and below the red line indicated in Fig.~\ref{fig:HD142527_IRDIS_PolBG_200pos_20avg_20pix}. We used the same inner and outer edge for both VBB (ZIMPOL) and $H$-band (IRDIS) observations to ensure that the profiles can be compared between both wavelengths. The outer boundary is therefore limited by the VBB data because the FOV of ZIMPOL is much smaller than the FOV of IRDIS. This kind of normalization of the radial profiles will lead to a small ($\sim 5 \%$) underestimation of the radially integrated polarized intensity in a given profile, but because we applied the same method to the intensity profiles $\tilde{I}_{\rm disk}(\varphi, r)$ as well, it will provide a high accuracy for the degree of polarization.

\subsubsection{Disk intensity profiles}
For the measurement of $\tilde{I}_{\rm disk}(\varphi, r)$, we first extracted radial profiles from the observed Stokes $I(\varphi, r) = I_{\star}(\varphi, r) + \tilde{I}_{\rm disk}(\varphi, r)$, following the same strategy as explained above for $Q_{\phi}$, and then subtracted from it the signal $I_{\star}(\varphi, r)$ of the bright stellar PSF. To simulate the stellar PSF profile $I_{\star}(\varphi, r)$, we applied a special form of RDI, using the reference star PSF images for ZIMPOL VBB and IRDIS $H$-band discussed in Sec.~\ref{sec:ZIMPOL observations}. This method proved to deliver reliable results for $\tilde{I}_{\rm disk}(\varphi, r)$ for both datasets, and it is a straightforward way to quantify the systematic errors of the measurement. From the VBB reference dataset we selected 120x1.1~sec exposures distributed over the whole observing run. From the $H$-band dataset we selected all 172x16~sec exposures. This ensured that both sets of reference images contained PSFs with a range of different observing conditions and AO performances. From each individual PSF we extracted 72 reference profiles $I_{\star, \rm i}^{\rm ref}(r)$ at position angles $\varphi=0^{\circ}, 5^{\circ}, 10^{\circ},\text{etc.}$, with each $I_{\star, \rm i}^{\rm ref}(r)$ being an azimuthal average over a $10^{\circ}$ section of the image. This resulted in a total of 8\,640 and 12\,384 different reference profiles for VBB and $H$-band, respectively. We then used a least-squares algorithm to fit each $I_{\star, \rm i}^{\rm ref}(r)$ to the $I(\varphi, r)$ profiles of each investigated dataset for HD142527. The only free parameter for the fit was the scaling factor $k_{\rm i}$. In Fig.~\ref{fig:HD142527_PSFbestfits_200pos} we show examples of several different $k_{\rm i} I_{\star, \rm i}^{\rm ref}(r)$ reference profiles compared to the radial profiles $I(\varphi, r)$ at position angle $\varphi = 200^{\circ}$ for the VBB and $H$-band. The cross-hatched areas highlight where the $I_{\star, \rm i}^{\rm ref}(r)$ were fit to $I_{\rm total}(\varphi, r)$ to determine $k_{\rm i}$. The position of the areas used for the fit were chosen for each radial profile individually such that they do not overlap significantly with the disk intensity. We used the mean of the $N=4000$ best fitting $k_{\rm i} I_{\star, \rm i}^{\rm ref}(r)$ profiles (red line) as approximation for $I_{\star}(r)$,

\begin{equation}
I_{\star}(r) \approx \frac{1}{N} \sum_{i=1} ^{N} k_{\rm i} I_{\star, \rm i}^{\rm ref}(r) = \langle k I_{\star}^{\rm ref}(r) \rangle.
\label{eq: make reference profiles}
\end{equation}

The quality of the fit was assessed through the root-mean-square (RMS) of the residuals after the fit, and the profiles with the smallest RMS were selected for Eq.~\eqref{eq: make reference profiles}. In addition, $k_{\rm i} I_{\star, \rm i}^{\rm ref}(r)$ with small RMS but $k_{\rm i}$ that deviated significantly from the expected scaling factor\footnote{Estimated through the difference in brightness difference between HD142527 and the observed reference star and the different exposure times.} were excluded from the calculation of the mean because their radial profiles can deviate significantly from the true $I_{\star}(r)$ between the cross-hatched areas.

In the $H$-band data, the mean reference profile models fit the increased intensity produced by the AO control ring at $\sim$~$0.9\arcsec$ very well (Fig.~\ref{fig:HD142527_PSFbestfits_200pos}, bottom). The second bump in the radial profile at $\sim 1.1\arcsec$ is caused by the disk intensity. In the VBB data, the disk is far outside the AO control ring, therefore the disk signal is less disturbed by the PSF bump caused by the AO control radius, but the brightness of the disk relative to the stellar PSF is much lower than in the $H$-band data.

As a last step, just like for $\tilde{Q}_{\phi, \rm disk}(\varphi, r)$, we then also fitted a line through $\tilde{I}_{\rm disk}(\varphi, r)$ at the same reference points inside and outside of the disk (cross-hatched areas in Fig.~\ref{fig:HD142527_IRDIS_PolBG_200pos_20avg_20pix}), along which we set the radial profile to zero. Detailed examples for the resulting $\tilde{I}_{\rm disk}(\varphi, r)$ at $\varphi = 200^{\circ}$ are presented in Fig.~\ref{fig:HD142527_200pos_profiles}, and all the measured radial profiles can be found in Appendix~\ref{sec:Radial profiles}. The radial profiles $\tilde{Q}_{\phi, \rm disk}(\varphi, r)$ and $\tilde{I}_{\rm disk}(\varphi, r)$ for each position angle yield a radial profile of the degree of polarization $p_{\rm disk}(\varphi, r) = \tilde{Q}_{\phi, \rm disk}(\varphi, r) /\tilde{I}_{\rm disk}(\varphi, r)$ and an intensity-weighted average degree of polarization $p_{\rm disk}(\varphi)$, which is denoted for each position angle in Appendix~\ref{sec:Radial profiles}.

Because the IRDIS observations were optimized for the detection of the disk (e.g., coronagraphic long exposure time), the signal of the disk is much stronger in the $H$-band data. However, it is more challenging to reliably subtract the stellar PSF from the disk signal and determine $\tilde{I}_{\rm disk}(\varphi, r)$ in these
data because of the AO control ring and the diffraction pattern of the telescope spiders (see Fig.~\ref{fig:HD142527_stokes_I_Qphi}). The diffraction spiders are present in the VBB data as well, but they are weaker and more diluted in the reduced data. The AO control ring is located at a distance of about 0.34" for the ZIMPOL VBB observations, which is clearly inside the inner rim of the disk, but about 0.84" in $H$-band, where it coincides with parts of the disk in HD142527. The southern part of the disk is farther away from the star than any other part (see Fig.~\ref{fig:HD142527_stokes_I_Qphi}, bottom right panel) and the AO control ring and the disk form two distinct radial features that can be separated most reliably as shown in Fig.~\ref{fig:HD142527_PSFbestfits_200pos} for a position angle of $200^{\circ}$. At other position angles, the disk signal and the AO control ring overlap, introducing larger uncertainties for the measured radial profiles, especially for the far side of the disk at about position angles $30^{\circ}$ to $90^{\circ}$. The narrow and bright surface brightness profiles on the near side of the disk generally allowed for more precise measurements of $\tilde{I}_{\rm disk}(\varphi, r)$ and eventually $p_{\rm disk}(\varphi, r)$. For several position angles, especially in the $H$-band data, measurements were impossible due to the bright diffraction spiders of the telescope. Some of these problems could be mitigated by performing measurements only on the 8~sec or 16~sec DIT data separately because the spider features at a given position angle can be absent in one of the two sets.

\subsubsection{Measurement errors}

The statistical noise of the individual data points in the radial profiles is low (VBB data: $\sim 4\cdot10^{-3}~{\rm cts}$; $H$-band data: $\sim 0.3~{\rm cts}$) because the data are binned. In the VBB data, the detector read-out noise dominates the statistical noise because be the number of counts at the separation of the disk is low (see Fig.~\ref{fig:HD142527_PSFbestfits_200pos}, top). In the $H$-band, the noise is composed of photon noise, speckle noise, and systematic errors from the extraction of the ${I}_{\rm disk}$ values.

For the uncertainty of $\tilde{Q}_{\phi, \rm disk}(\varphi, r)$, we only considered the statistical errors because other noise sources are eliminated by the polarimetry. For $\tilde{I}_{\rm disk}(\varphi, r)$, the systematics from the measurement of the radial profile usually dominate the error budget. To calculate the systematic errors, we determined the 68.3\% spread of the 4000 best-fitting reference profiles $k_{\rm i} I_{\star, \rm i}^{\rm ref}(r)$. The final error bars of $\tilde{I}_{\rm disk}(\varphi, r)$ are then calculated by adding the statistical and systematic errors. The errors for the degree of polarization $p_{\rm disk}(\varphi, r)$ were calculated with error propagation from the errors of $\tilde{Q}_{\phi, \rm disk}(\varphi, r)$ and $\tilde{I}_{\rm disk}(\varphi, r)$.

\subsubsection{Integrated disk intensity and degree of polarization}
\label{sec:Total disk flux and fractional polarization}

Instead of using large-aperture photometry, the integrated disk intensities $Q_{\phi, \rm disk}$ and $I_{\rm disk}$ can also be calculated by integrating $\tilde{Q}_{\phi, \rm disk}(\varphi, r)$ and $\tilde{I}_{\rm disk}(\varphi, r)$  in radial and azimuthal direction and using $f_{\rm corr}^P$ and $f_{\rm corr}^I$ to correct for PSF smearing. However, as mentioned before, the normalization of the radial profiles can lead to an underestimation of the intensity values. Therefore the integrated radial profiles only provide a lower limit for the integrated disk intensity values,

\begin{equation}
Q_{\phi, \rm disk} > f_{\rm corr}^P \int_{0}^{2\pi} \int_{r_{min}(\varphi)}^{r_{max}(\varphi)} \tilde{Q}_{\phi, \rm disk}(\varphi, r)\,r\,dr\,d\varphi
\label{equ: qphi profile integration}
\end{equation}

\begin{equation}
I_{\rm disk} > f_{\rm corr}^I \int_{0}^{2\pi} \int_{r_{min}(\varphi)}^{r_{max}(\varphi)} \tilde{I}_{\rm disk}(\varphi, r)\,r\,dr\,d\varphi.
\label{equ: int profile integration}
\end{equation}

Because we measured $\tilde{Q}_{\phi, \rm disk}(\varphi, r)$ and $\tilde{I}_{\rm disk}(\varphi, r)$ only for a limited set of $\varphi$ and $r$ with some sector missing, we linearly interpolated the missing values in order to calculate the integral. The calculated lower limits for $Q_{\phi, \rm disk}/I_{\rm total}$ and $I_{\rm disk}/I_{\rm total}$ are $0.29 \%$ and $1.3 \%$ for the VBB data and $0.90 \%$ and $3.4 \%$ for the $H$-band, which is consistent with the measurements from the large-aperture photometry presented in Sec.~\ref{sec:Integrated disk flux analysis}.

More importantly, the derived profiles $Q_{\phi, \rm disk}$ and $I_{\rm disk}$ can be used to determine the degree of polarization $p_{\rm disk}$ with high precision because the normalization of the radial profiles affects the intensity and polarized intensity in the same way and therefore barely changes the degree of polarization (see Fig.~\ref{fig:HD142527_200pos_sim_profiles_Hband}). By integrating the radial profiles as shown in Eq.~(\ref{equ: qphi profile integration}) and (\ref{equ: int profile integration}), but separately from position angles $-20^{\circ}$ to $160^{\circ}$ and $160^{\circ}$ to $340^{\circ}$, it is also possible to calculate the degree of polarization just for the far side $p_{\rm disk}^{\rm far}$ and near side $p_{\rm disk}^{\rm near}$ of the disk. The results of these measurements are reported in Table~\ref{table: all relative measurements}.

\section{Results}
\label{sec:Results}
\subsection{Final photopolarimetric values}
\label{sec:precise measurements}
The most precise measurements described in Sec.~\ref{sec:Data analysis} are the integrated polarized intensity $Q_{\phi, \rm disk}$ in large apertures and the degree of polarization $p_{\rm disk}$ from the radial profiles. The final values are summarized in Table~\ref{table: all relative measurements}. The total disk intensity $I_{\rm disk}$ could not be measured precisely but was instead calculated from $Q_{\phi}$ and $p$. The brightness magnitudes and spectral flux densities for the total system flux were obtained by using the measured magnitude and flux values from the Tycho-2 \citep{Hog00}, GAIA \citep{GAIA18}, and 2MASS \citep{Cutri03} surveys to produce an SED of HD142527 from $B$- to $K$-band. We then used the GAIA GRP and 2MASS $H$-band filter transmission curves to correct for the different filter responses compared to the ZIMPOL VBB filter and the IRDIS $H$-band, and we used a blackbody spectrum at 9730~K to obtain the correct magnitudes in the Vega system. The errors reported for the magnitude and flux values in Table~\ref{table: all absolute measurements} result from the combination of our own measurement errors, the measurement errors of GAIA and 2MASS, the variability of HD142527 determined with Hipparcos, and the estimated interpolation errors caused by the low spectral resolution of the SED that was used for HD142527. The magnitudes derived for HD142527 with the ZIMPOL VBB filter and the IRDIS $H$-band filter are within $\pm 0.10$ and $\rm 0.05$ of the values measured with the corresponding GAIA and 2MASS filters, respectively.

\begin{table*}
\caption{Relative photopolarimetric measurements for HD142527 from this work}
\label{table: all relative measurements}
\centering
\begin{tabular}{l|lll|lll}
\hline
 & \multicolumn{3}{c}{VBB (0.735~$\mu$m)} & \multicolumn{3}{c}{$H$ (1.625~$\mu$m)} \\
 & total & near & far & total & near & far \\
\hline
$Q_{\phi, \rm disk}/I_{\rm total}$ [\%] & $0.46\pm0.02$ & $0.23\pm0.01$ & $0.23\pm0.01$ & $1.07\pm0.07$ & $0.51\pm0.04$ & $0.56\pm0.05$ \\
$p_{\rm disk}$ [\%] & $21.8\pm0.4$ & $17.1\pm0.4$ & $28.0\pm0.9$ & $26.7\pm0.4$ & $20.2\pm0.6$ & $35.1\pm2.1$ \\
$I_{\rm disk}/I_{\rm total}$ [\%] ~\tablefootmark{a} & $2.1\pm0.1$ & $1.3\pm0.1$ & $0.82\pm0.06$ & $4.0\pm0.3$ & $2.5\pm0.3$ & $1.6\pm0.2$ \\
\hline
\end{tabular}
\tablefoot{\tablefoottext{a}{The values for $I_{\rm disk}$ were not measured directly from the data, but were calculated from $Q_{\phi, \rm disk}/p_{\rm disk}$.}}
\end{table*}

\begin{table}
\caption{Absolute flux values for HD142527 from this work}
\label{table: all absolute measurements}
\centering
\begin{tabular}{llll}
\hline
Band & $I_{\rm total}$ & $I_{\rm disk}$ & $Q_{\phi, \rm disk}$ \\
\hline
\multicolumn{3}{l}{$\left( {\rm mag} \right)$} & \\
VBB & $7.62\pm0.03$ & $11.81\pm0.03$ & $13.46\pm0.03$ \\
$H$ & $5.76\pm0.07$ & $9.25\pm0.07$ & $10.69\pm0.07$ \\
\hline
\multicolumn{3}{l}{$\left( {\rm erg/s/cm^2/\mu m}  \cdot 10^{-10} \right)$} & \\
VBB & $136\pm4$ & $2.86\pm0.08$ & $0.626\pm0.018$ \\
$H$ & $58.5\pm3.5$ & $2.34\pm0.14$ & $0.626\pm0.037$ \\
\hline
\multicolumn{3}{l}{$\left( {\rm mJy} \right)$ ~\tablefootmark{a}} & \\
VBB & $2450\pm70$ & $51.4\pm1.5$ & $11.3\pm0.3$ \\
$H$ & $5150\pm310$ & $206\pm12$ & $55.1\pm3.3$ \\
\hline
\end{tabular}
\tablefoot{\tablefoottext{a}{The values for the spectral flux density in [${\rm erg/s/cm^2/\mu m}$] were converted into mJy by using the central wavelength of the filter from the corresponding observations.}}
\end{table}

\subsection{Brightness and degree of polarization depending on position angle}
\begin{figure*}
\centering
\begin{tabular}{ll}
a) & b) \\
\includegraphics[width=8.6cm]{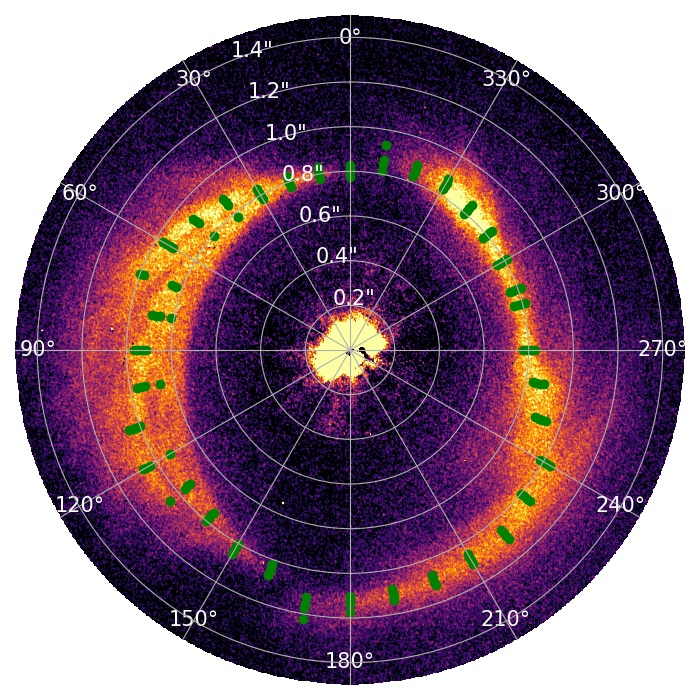} &  \includegraphics[width=8.6cm]{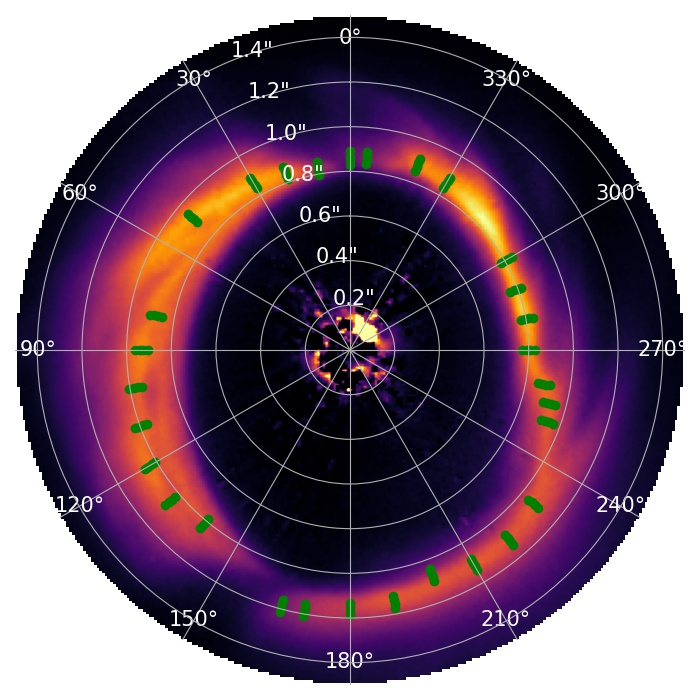}\\
c) & d) \\
\includegraphics[width=8.6cm]{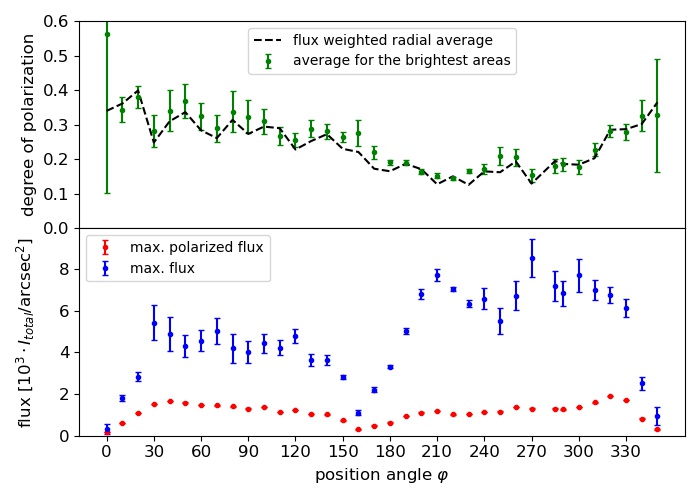} & \includegraphics[width=8.6cm]{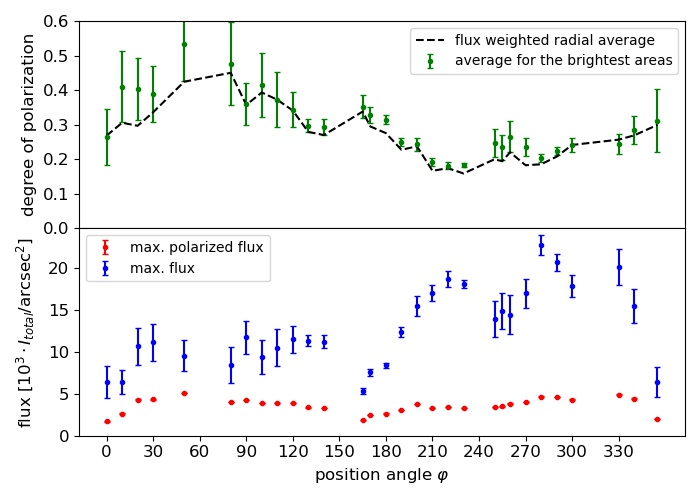} \\
\end{tabular}
\caption{Polarized intensity of HD142527 in the VBB (a) and $H$-band (b) and the corresponding surface brightness $\tilde{I}_{\rm disk}(\varphi)$, maximum polarized surface brightness $\tilde{Q}_{\phi, \rm disk}(\varphi),$ and degree of polarization $p_{\rm disk}(\varphi)$ in (c) and (d), respectively. The azimuthal surface brightness profiles in (c) and (d) show the average surface brightness for the four brightest values (indicated by the green dots in (a) and (b)) in each radial profile (see Appendix~\ref{sec:Radial profiles}). The values are not corrected for signal loss due to PSF smearing. The plots for the azimuthal degree of polarization also show the degree of polarization at the positions where the highest surface brightness was measured (green dots), and additionally show the intensity-weighted average polarization for the complete radial profiles (dashed line).}
\label{fig:ang_dep_frac_pol}
\end{figure*}

The radial profiles characterize the radial and azimuthal variation of the disk surface brightness and $p_{\rm disk}$. The measurements summarized in Fig.~\ref{fig:ang_dep_frac_pol} show for each position angle $\varphi$ the $\tilde{Q}_{\phi, \rm disk}(\varphi, r)$, $\tilde{I}_{\rm disk}(\varphi, r)$ and $p_{\rm disk}(\varphi, r)$ values from the four brightest points of the $\tilde{Q}_{\phi, \rm disk}(\varphi, r)$ radial profiles from Appendix~\ref{sec:Radial profiles} (in other words, measured along the bright ridge). The location of the brightest points for each position angle are marked in panels a) and b). We chose not to display the average or integrated total intensity for each position angle because these values depend on our selection of the inner and outer edge of the radial profiles, which had to consider the FOV of ZIMPOL, faint disk sections, and PSF noise features. The brightest points, however, show little dependence on this empirical choice and therefore provide reliable measurements of the disk surface brightness. These points also provide the most precise values of $p_{\rm disk}(\varphi, r)$. However, because $p_{\rm disk}$ is a normalized quantity, it can also be calculated as the intensity-weighted average over the whole radial profile as long as it is not strongly dependent on $r$. We show this by comparing the intensity-weighted radial average $p_{\rm disk}(\varphi)$ (dashed black line) versus the intensity-weighted average of the brightest points (green dots). The values agree very well at most position angles, except for the $H$-band values in the range between $10^\circ$ to $80^\circ$. These values are affected by systematic noise from the AO control ring, which seems to cause an underestimation of the brightness from the brightest spots on the disk.

The azimuthal profiles are similar for both wavelengths and clearly show the strong $\tilde{I}_{\rm disk}(\varphi)$ asymmetry between the near and far sides of the disk, while $\tilde{Q}_{\phi, \rm disk}(\varphi)$ is symmetrical. This produces the significant asymmetry of $p_{\rm disk}(\varphi)$ with a broad $\Delta \varphi \approx 50^{\circ}$ minimum and maximum degree of polarization centered on the expected location of the semiminor axis of the disk at a position angle of $70^{\circ} / 250^{\circ}$.

The surface brightness in Fig.~\ref{fig:ang_dep_frac_pol} is given relative to the star, but
the stellar magnitudes from Table~\ref{table: all absolute measurements} allow a simple transformation into magnitudes per arcsecond squared with 

\begin{equation}
m_{\rm disk} = m_{\star} - 2.5\cdot\log\left(f_{\rm corr}^I \tilde{I}_{\rm disk}\right).
\label{equ: magnitude transformation}
\end{equation}

This formula includes the PSF smearing correction factors $f_{\rm corr}^I$ for the integrated radial profiles (Table~\ref{table: PSF smearing correction factors}). The correction factors are precise within the error bars stated in Table~\ref{table: PSF smearing correction factors} for the radially integrated intensity values, but not necessarily for individual points on the radial profiles. Especially the maximum intensity of narrow peaks, as observed at the disk rim on the near side, is more strongly effected by PSF smearing and would require larger correction factors (up to 50\% larger). The accurate calculation of an unconvolved maximum surface brightness on the near side would therefore require deriving individual correction factors for each point on the radial profile. Therefore we limit the analysis here to the far side of the disk at about a position angle of $70^{\circ}$, where we determine a more accurate PSF smearing corrected surface brightness with the factors in Table~\ref{table: PSF smearing correction factors} for all individual points on the radial profiles because the profiles are relatively flat. With Eq.~(\ref{equ: magnitude transformation}) we calculate a surface brightness of $12.9\pm0.2~{\rm mag/arcsec^2}$ in VBB and $10.2\pm0.2~{\rm mag/arcsec^2}$ in $H$-band for the points marked in Fig.~\ref{fig:ang_dep_frac_pol}. This translates into a spectral flux density of $18\pm3~{\rm mJy/arcsec^2}$ in VBB and $87\pm16~{\rm mJy/arcsec^2}$ in $H$-band. The results in $H$-band are on the same order as the results presented in \citet{Honda09}, but a detailed comparison is not possible because the authors did not investigate the effect of PSF smearing in their work and only analyzed a small section of the near side of the disk around the position angle $240^{\circ}$.

\subsection{Simulating near- and far-side brightness profiles}
\label{sec:Near- and far-side intensity profiles}
\begin{figure*}
\centering
\begin{tabular}{c}
\includegraphics[width=\linewidth]{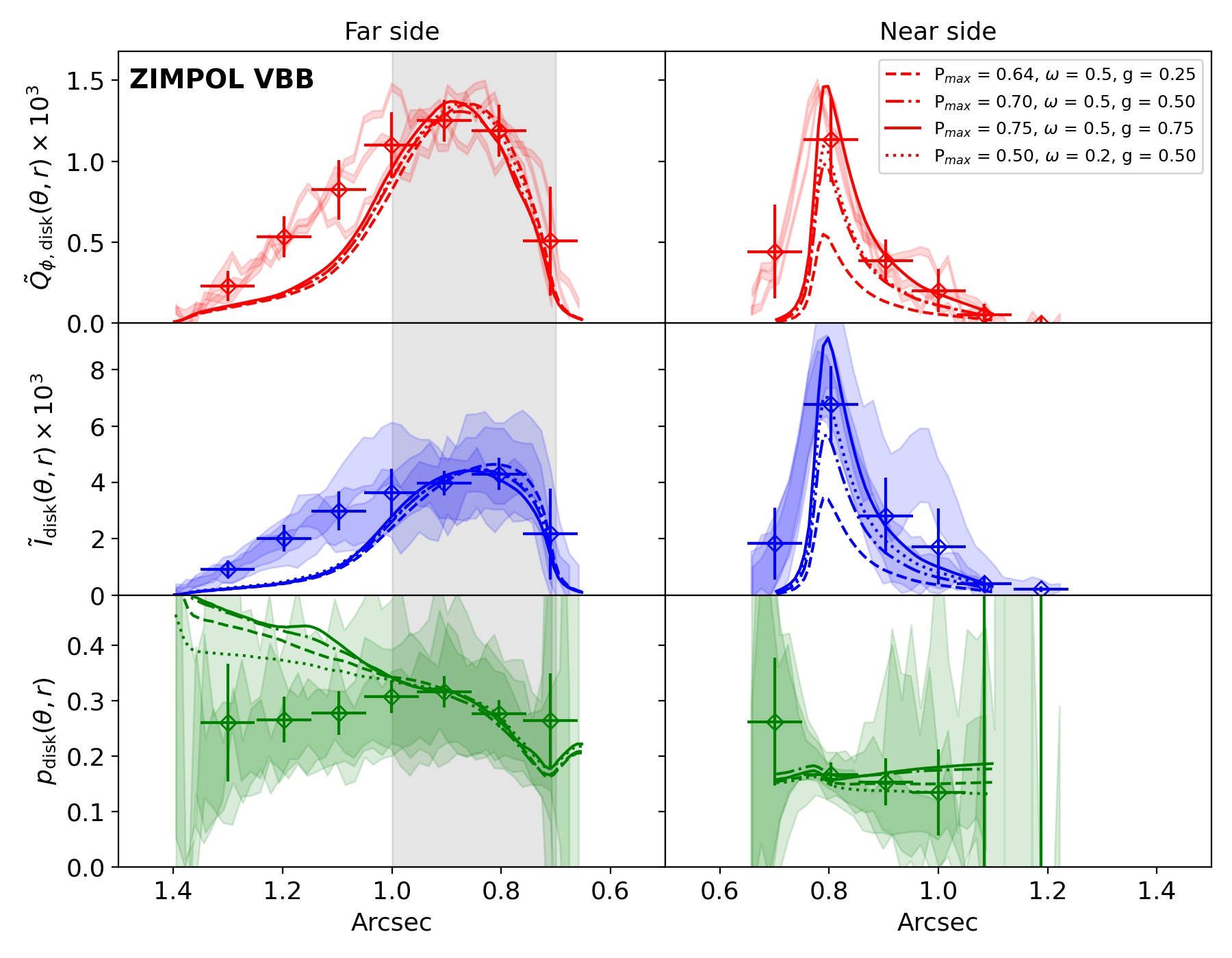} \\
\end{tabular}
\caption{Comparison of PSF-convolved model calculation for the radial profiles of $\tilde{Q}_{\phi, \rm disk}$, $\tilde{I}_{\rm disk}$ , and $p_{\rm disk}$ with measured profiles along the minor axis of HD142527. The lines show simulation results for different dust-scattering parameter combinations for the geometry plotted in Fig.~\ref{fig:tau_0_01_surface}. The shaded bands show a selection of three measured radial profiles close to the semiminor axes of the disk on the far and near side of the disk at position angles of $70^{\circ}$, $80^{\circ}$, and $90^{\circ}$ and $270^{\circ}$, $285^{\circ}$, and $290^{\circ}$, respectively, in units of surface brightness relative to the total flux of the system, i.e., [$I_{\rm total}/{\rm arcsec}^2$], and the diamonds show the average and the spread of all data points for all radial profiles combined. The model profiles are normalized and fitted to the measurements between $0.7\arcsec$ and $1\arcsec$ on the far side (shaded gray regions), where the plane-parallel surface approximation applies best.}
\label{fig:near_far_profiles}
\end{figure*}

\begin{figure}
\begin{tabular}{c}
\includegraphics[width=\linewidth]{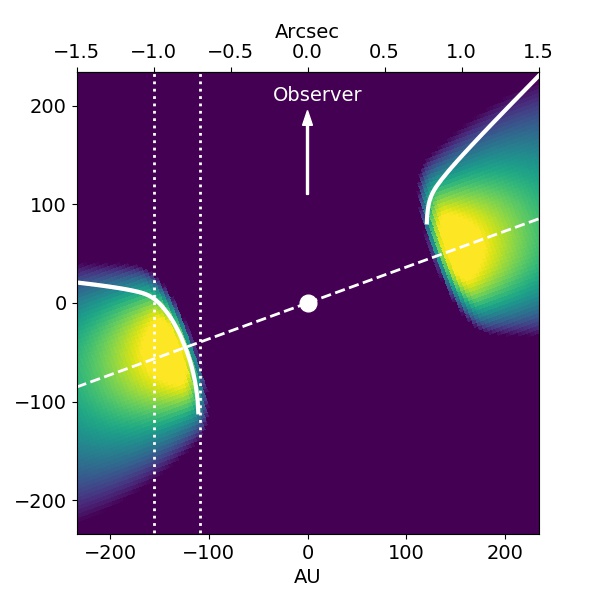} \\
\end{tabular}
\caption{Disk model geometry for HD142527 for the vertical plane along the minor axis with the location of the visible part of the reflecting disk surface (solid white line). The color scale illustrates the adapted disk volume density model of \citet{Marino15}. The disk midplane (dashed white line) is inclined by $20^{\circ}$ toward the observer. The dotted lines indicate the location of the steep wall on the far side of the disk between $0.7\arcsec$ and $1.0\arcsec$ where the approximation of the disk wall by a plane-parallel surface is expected to be quite good.}
\label{fig:tau_0_01_surface}
\end{figure}

The near- to far-side asymmetry of the observed disk surface brightness and degree of polarization can also be seen in the individual $\tilde{Q}_{\phi, \rm disk}(r, \varphi)$, $\tilde{I}_{\rm disk}(r, \varphi)$ and $p_{\rm disk}(r, \varphi)$ radial profiles (Fig.~\ref{fig:HD142527_Hband_radial_profiles}). We illustrate this in Fig.~\ref{fig:near_far_profiles} with a few selected radial profiles located close to the semiminor axis on the near and far side. The polarized surface brightness profiles $\tilde{Q}_{\phi, \rm disk}(r, \varphi)$ show similar maximum values on both sides, while the maximum values of $\tilde{I}_{\rm disk}(r, \varphi)$ are significantly higher on the near side compared to the far side. Therefore the degree of polarization is significantly lower on the near side. Both effects can be explained by the difference in scattering angle and the presence of forward-scattering dust. The scattering angle for the far side decreases with separation and lies between $\approx 120^\circ$ and $90^{\circ}$ and the produced degree of polarization is high, while the scattering angle on the near side is about $50^\circ$, which produces a lower degree of polarization.

We constrained the optical properties of the dust with simulations of the photon absorption and scattering with Monte Carlo simulations of the photon random walk (Ma \& Schmid, in prep.). The photons are sent out by the central star and undergo single and multiple scatterings or absorption in the disk wall. The angle dependence of the scattering intensity is calculated with the Henyey-Greenstein phase function, and the angle dependence of the degree of  polarization is modeled in the same way as for Rayleigh scattering. The disk wall is approximated locally by a plane-parallel model, and the scattered photons escape from the same position as they penetrated the wall. The optical properties of the dust are described by the asymmetry parameter $g$, the maximum polarization $P_{\rm max}$ for a scattering angle of $90^\circ$, and the single-scattering albedo $\omega$. We simulated the reflected light from the disk for the far side $\varphi = 70^{\circ}$ and the near side $\varphi = 250^{\circ}$ adopting the geometry of the disk surface based on the axisymmetric disk model for HD142527 from \cite{Marino15}, which provides a detailed dust density distribution in radial and vertical direction. We also adopted a surface of constant optical depth to define the disk wall geometry. To improve the match with our observations, we changed the length scales of the disk to account for the new GAIA distance to HD142527 of 156~pc instead of 140~pc, adopted a disk inclination of $20^\circ$, and moved the near side farther out radially (by $\sim$10~AU) to account for the quite substantial deviations from axisymmetry of HD142527. A vertical cut through our inclined disk model is shown in Fig.~\ref{fig:tau_0_01_surface}, illustrating the visible parts of the adopted disk surface geometry for the Monte Carlo simulations and the underlying density distribution model from \cite{Marino15}.

The approximation of the disk wall by a plane-parallel surface is expected to be quite good for the central section of the wall on the far side, which is close to perpendicular to the midplane. The strongly curved rims of the disk walls and the flat upper disk surfaces are more problematic; a plane-parallel geometry is probably a poor approximation here. Therefore we fit our model results to the measured profile sections for the steep wall on the far side between $0.7\arcsec$ and $1.0\arcsec$ and investigated how well the model profile matches other disk regions. For our models we first selected the model scattering parameters $\omega$ and $g$ for the scattering albedo and asymmetry and searched for a $P_{\rm max}$ value that provided a good match with the measured degree of polarization of the wall on the far side. To first order, the simulated $p_{\rm disk}$ depends linearly on $P_{\rm max}$ for given $\omega$ and $g$. This yields a family of scattering parameters that give the correct degree of polarization $p_{\rm disk}$ for the wall on the far side. The modeled surface brightness levels of the calculated $\tilde{I}_{\rm disk}$ were scaled to fit the measured brightness on the wall of the far side. We discuss in Section~\ref{sec: Disk albedo} constraints on this scaling based on estimates for the stellar illumination and the disk surface reflectivity.

A few examples of simulated and PSF convolved radial profiles for the disk minor axis with scattering parameters that match the polarization level $p_{\rm disk}$ on the far side well are shown in Fig.~\ref{fig:near_far_profiles} in comparison with the measured disk profiles. As in the observations, the reflected intensity on the far side around $r=0.7\arcsec$ first increases faster than the polarized intensity with a corresponding $p_{\rm Disk}(r)$ dependence of 0.25 at 0.7\arcsec and 0.31 at 0.9\arcsec. Other sections of the simulated profiles show some shortcomings that are most likely caused by the adopted plane-parallel surface geometry. For example, at larger separations on the far side ($>1\arcsec$), the simulated brightness drops faster with separation for all parameter combinations. In addition, the modeled degree of polarization increases rather than roughly remaining at a constant level.

The simulated radial profiles on the near side show a strong dependence on the asymmetry parameter $g,$ and a good match is obtained for $g = 0.75$. However, we expect that the plane-parallel approximation overestimates the real $g$ factor  because the real HD142527 disk will produce a significant contribution to the forward scattering peak from optically thin dust layers located above the near side disk rim and the upper disk surface (see Fig.~\ref{fig:tau_0_01_surface}). A flat surface model can compensate for this problem by boosting the forward-scattering with a large $g$ parameter.

We find the following dependences and constraints for the three dust scattering parameters: The observed degree of polarization $p_{\rm disk}$ is directly proportional to the maximum polarization $P_{\rm max}$, therefore $P_{\rm max}$ must be larger than the observed maximum $p_{\rm disk}$ because of depolarization from multiple scattering, which defines the strong constraints $P_{\rm max}\approx30~\%$ for the VBB and $P_{\rm max}\approx40~\%$ for the $H$-band. Increasing the single-scattering albedo $\omega$ enhances the amount of reflected light, but it also reduces $p_{\rm disk}$ because multiply scattered photons with randomized polarization contribute to the reflected light, but dilute the polarization signal. Therefore we only obtain a good match for the polarization of the wall on the far side if both $\omega$ and $P_{\rm max}$ are high (e.g., $\omega=0.5, P_{\rm max}=0.70$), or both are low (e.g., $\omega=0.2, P_{\rm max}=0.50$), with some lower dependences on the $g$ parameter as well. Our tests have also shown that this relation results in an upper limit $\omega<0.7$ because even for $P_{\rm max}=1$, $\omega$ should not exceed 0.7 to match the data. Increasing the asymmetry parameter $g$ mainly changes the ratio of the reflected intensity of the near and far sides of the disk (see Fig.~\ref{fig:near_far_profiles}). Unfortunately, our surface models are not suited to accurately model the near side of the disk, and therefore our constraint $g<0.75$ is rather loose. Other than that, the $g$ parameter also affects the degree of polarization somewhat, and strong forward-scattering or high $g$ values increase the probability that scattered photons undergo additional interactions in the disk. This results in an increased probability for multiple scattering and reduced $p_{\rm disk}$.

\subsection{Disk reflectivity}
\label{sec: Disk albedo}
The observed disk surface brightness shown in Fig.~\ref{fig:near_far_profiles} is expressed relative to the total stellar intensity as the surface brightness contrast ${\rm SB}/I_\star$. The peak value at the wall on the far side in the VBB is ${\rm SB}/I_\star=6.7\cdot 10^{-3} {\rm arcsec}^{-2}$ if corrected for the PSF smearing (correction factor of 1.67 from Table~\ref{table: PSF smearing correction factors}), which results in an intrinsic magnitude difference ${\rm SB}~[{\rm mag\,arcsec}^{-2}] - m_\star [{\rm mag}]=5.4~{\rm mag/arcsec}^2$. This surface brightness contrast can be used to estimate the surface reflectivity $R$ of the wall on the far side, which is defined relative to an ideal white Lambert surface according to $R = 1$ for $R=R_{\rm Lam}$. The reflectivity per unit area of a Lambert surface is $R_{\rm Lam}(\theta)=\cos\theta/\pi,$ where $\theta$ is the
viewing angle with respect to the surface normal. The disk surface brightness is measured per unit area perpendicular to the the viewing direction, or $R_{\rm SB,Lam}=1/\pi$, and it must consider the stellar illumination, 
\begin{equation}
  {\rm SB} = R \,\frac{L_\lambda \cos\theta_0}{4\pi\,d^2}\,\frac{1}{\pi}\,,
\end{equation}
where $d$ is the physical distance of $\approx 120-150$~AU, and $\theta_0$ is the incidence angle $\approx 0^\circ - 45^\circ$ (or $\cos\theta_0\approx 1 - 0.7$) for the wall on the far side (see Fig.~\ref{fig:tau_0_01_surface}). Expressing $d$ in arcsec yields a simple relation between the surface brightness contrast per arcsec$^2$ and the surface reflectivity,
\begin{equation}
R =\frac{{\rm SB}}{I_\star}\,\cdot\,\frac{\pi}{\cos\theta_0}(d[{\rm arcsec}])^2\,.
\end{equation}  
This relation assumes that stellar emission is isotropic and that neither the observed stellar intensity $I_\star=L_\lambda/(4\pi\,D^2)$ nor the disk irradiation $I_0 =  L_\lambda/(4\pi\,d^2)$ are strongly affected by absoption or additional emission. This assumption is probably not well fulfilled for HD142527 because observations show scattering emission \citep{Avenhaus17}, thermal emission \citep{Verhoeff11}, and absorption \citep{Marino15} by hot dust located close $<5$~AU to the star. Therefore we expect uncertainties of at least $\pm 30~\%$ for the deduced reflectivity.

According to the disk surface model (Fig.~\ref{fig:tau_0_01_surface}), the parameters for the wall on the far side near the disk midplane are about $\cos \theta_0=1$ and $d=0.8\arcsec$ or $R = 0.015\pm0.005$. For the $H$-band, the convolved surface brightness contrast is roughly 2.2 times higher (see Fig.~\ref{fig:ang_dep_frac_pol}), but the correction factor for PSF smearing is lower. This results in an $H$-band reflectivity of $0.03\pm0.01$. 

This low reflectivity requires that either the single-scattering albedo $\omega$ is low or that the scattering asymmetry $g$ is very high, so that most photons are scattered farther into the disk where absorption can occur \citep{Mulders13}. For example, the simulated reflectivity for the wall on the far side near the midplane is about $R=0.018$ for the parameters $\omega=0.2,\, g=0.5,$ or $R=0.03$ for $\omega=0.5 \text{ and}\, g=0.75$ according to calculations for plane-parallel surfaces (Ma \& Schmid, in prep.), and these two cases appear to approximately bracket the good values for the dust in HD142527. We can also say that these model parameters would then suggest a maximum scattering polarization of $P_{\rm max}\approx 0.5$ for the $\omega=0.2$ case (VBB) and perhaps $P_{\rm max} \approx 0.7$ for $\omega=0.5$ for the $H$-band case.

\section{Discussion}
Differential polarimetric imaging of protoplanetary disks has seen steady improvements in the past decade from disk detection to high-contrast disk characterization with small inner working angles $<0.1\arcsec$ . With the newest generation of instruments, now a level is also reached at which quantitative photopolarimetric measurements become possible \citep{Perrin15,Schmid18,deBoer20,vanHolstein20}. This provides the scattering properties of the circumstellar dust and helps improve our understanding of dust evolution and the formation of planets in these disks.

Accurate polarization parameters for circumstellar disks are not easy to
derive with high-contrast observations using extreme AO systems
because of the strongly variable PSF. Knowing the PSF well is essential for calibrating the extended polarized emission of the disk. Moreover, it is often difficult or even impossible to disentangle the disk intensity signal from the very strong and variable stellar PSF. Even the differential polarization signal $Q_\phi$ is very useful, but many ambiguities in the photopolarimetric analysis are solved when the degree of polarization $p_\phi=Q_\phi/I$ for the scattered disk radiation can also be derived. Only a few measurements available to date, and according to our knowledge, no measurements except for this work that include a detailed assessment of the uncertainties.

The selection of a favorable disk target under these circumstances is crucial to achieve accurate photopolarimetric measurements. HD142527 is an ideal target because the bright, strongly illuminated inner disk wall is located at a large apparent separation of $0.8\arcsec-1.1\arcsec$ , and therefore it is relatively easy to separate the disk from the bright star. In addition, PSF smearing and polarization cancellation effects are much smaller for extended disks. However, we caution that the disk of HD142527 is a special case, in which hot dust near the star produces a strong near-IR excess, shadows on the outer disk, and observable strong scattering very close to the star. For this reason, it is unclear how well the observed stellar intensity $I_*$ represents the illumination for the disk region. 

\begin{figure}
\includegraphics[width=\linewidth]{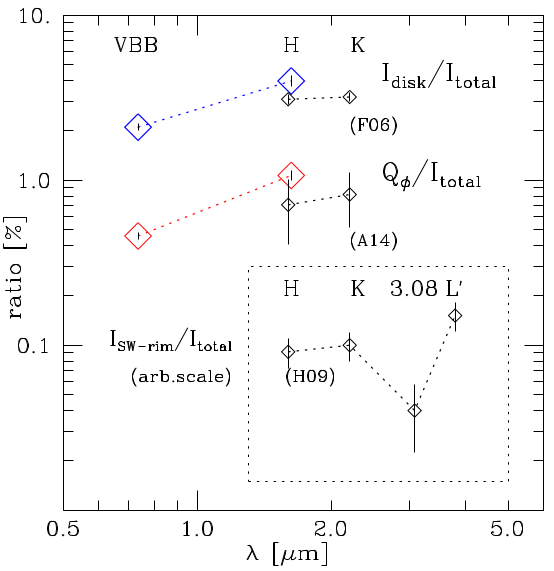} \\
\caption{Wavelength dependence of the measured relative intensities $I_{\rm disk}/I_{\rm total}$ and polarized intensities $Q_{\phi,{\rm disk}}/I_{\rm total}$ for the scattered light from the disk in HD142527. The colored symbols are from Table~\ref{table: all relative measurements}, the values marked with (F06) from \citet{Fukagawa06} and (A14) from \citet{Avenhaus14}. The values (H09) from \citet{Honda09} are given in an inset with an arbitrary scale factor to offset them from the other data because these are relative measurements for an area at the southwest rim of the disk.}
\label{fig:Dustcolor}
\end{figure}

\subsection{Color of the scattered light in HD142527}
The main results of this study are the photopolarimetric measurements for the VBB and the $H$-bands for the scattered light from the disk given in Table~\ref{table: all relative measurements}. The large wavelength separation of the two bands is ideally suited to quantify the spectral dependence of the scattered radiation. Figure~\ref{fig:Dustcolor} shows our results together with previous measurements found in the literature. We find a red color for the polarized intensity $Q_{\phi,{\rm disk}}/I_{\rm total}$ and the intensity $I_{\rm disk}/I_{\rm total}$ for the normalized
scattered radiation from the disk in HD142527. The measured values are about a factor of two higher for the $H$-band (1.625~$\mu$m) when compared to the VBB (735~nm) or when the relative intensities are roughly proportional to the wavelength.

For the disk-averaged degree of polarization, we find values of $p_{\rm disk}=21.8~\%$ in the VBB and $26.7~\%$ in the $H$-band. The polarization $p_{\rm disk}$ strongly depends on the disk azimuth angle. For both filters, $p_{\rm disk}$ is about 1.3 times higher for the far side and 0.77 times lower for the near side than the averaged value. This difference is caused by the large asymmetry of the disk intensity, while the polarized intensity is essentially the same for the near and far sides at both wavelengths. The maximum disk polarization ${\rm max}(p_{\rm disk})$ measured for the brightest disk regions on the far side are about $30\pm5~\%$ in the VBB and $40\pm10~\%$ in the $H$-band (see Fig.~\ref{fig:ang_dep_frac_pol}).

We also estimate a surface reflectivity of the wall on the far side and obtain $R=1.5~\%$ for VBB and $3.0~\%$ for the $H$-band. Based on very simple calculations for a plane surface, we conclude that these reflectivity values can only be explained by strongly forward-scattering dust $g\approx 0.5-0.75$ with a low single-scattering albedo $\omega \approx 0.2-0.5$. These estimates are based on the assumption that the stellar emission is spherically symmetric or that the disk "sees" the same stellar luminosity as an observer on Earth. This assumption can be questioned for the HD142527 system, which shows strong shadows from a close-in circumstellar disk \citep{Marino15} and a very strong IR excess of $F_{IR}/F_* = 0.92$ \citep{Dominik03} from thermal dust emission. This suggests that the light from the central star could be attenuated differently for our line of sight and for the illuminated far side of the disk. If the stellar light is strongly attenuated along our line of sight, then the real disk surface reflectivity would be even lower than 2 to 3\%\ , but the relatively strong IR excess could be easily explained. If there is strong absorption between the star and the far side of the disk (but not between the star and us), then the reflectivity would be higher, but the high relative IR excess would be harder to explain. This discussion is important because it questions the origin of the observed color and  the wavelength-dependent reflectivity. The observed color and reflectivity of the disk could at least partially be caused by wavelength-depended anisotropic emission from the central object. A comprehensive study of this potential effect is beyond the scope of this paper, however. The effect of anisotropic emission in systems with complex close-in dust structures on measurements of the reflected light should be investigated further in the future.

\subsection{Comparison with previous HD142527 data}

For the first time, we have presented scattered-light intensities for the disk around HD142527 for visual wavelengths ($< 1~\mu$m). However, several previous measurements in the IR can be compared to our $H$-band results. They are included in Fig.~\ref{fig:Dustcolor}. \citet{Fukagawa06} measured the relative intensity $I_{\rm disk}/I_{\rm total}$ of the scattered light for the $H$-band. Their result is 22~\% lower than our value. This can be considered as good agreement considering that the two studies used different apertures for the disk intensity measurement, and the observations of \citet{Fukagawa06} were performed at a much lower Strehl ratio. \citet{Fukagawa06} also measured the $K$-band intensity of the disk and found essentially the same $I_{\rm disk}/I_{\rm total}$ ratio as for the $H$-band, indicating that the $H-K$ color of the scattered light is gray. It would be interesting to obtain measurements for more bands between the VBB and $H$-bands for HD142527 to determine the wavelength of the turnover from red to gray colors for the scattered intensity. Scattered-light intensity measurements were also obtained by \citet{Honda09} in the $H$- and $K$-bands, the 3.08~$\mu$m ice absorption filter, and the $L'$-band (3.8~$\mu$m) for a bright region on the near side (southwest rim) of the disk. These data do not provide a disk-integrated intensity, but relative brightness measurements for the selected disk areas. They confirm the gray color for the $H$- and $K$-bands of \citet{Fukagawa06}, and report slightly red colors for $K-L'$ and a very strong attenuation of the scattered-light intensity in the 3.08~$\mu$m filter, most likely because icy grains are present in the scattering region.

The polarized intensity $Q_{\phi,{\rm disk}}/I_{\rm total}$ was determined previously by \citet{Avenhaus14} for the $H$- and $K$-bands with data from NACO/VLT. They obtained the same disk morphology as observed by us, with about equal polarized intensity on the east and west side and the same intensity asymmetry. They measured $Q_{\phi,{\rm disk}}/I_{\rm total}=0.71~\%$ for the $H$-band, or about one-third less when compared to our work. This result is compatible with this study because \citet{Avenhaus14} did not consider polarimetric cancellation effects and expected systematic uncertainties of up to 30~\%. Their $H-K$ color for the polarized intensity is quite uncertain (see Fig.~\ref{fig:Dustcolor}), but also compatible with gray dust scattering in the IR.

The degree of polarization $p_{\rm disk}$ was also derived by \cite{Avenhaus14} for the $H$-band. They measured $20~\%$ on the bright west side and about $45~\%$ with maxima above $50~\%$ on the east side. This agrees well with our results (Fig.~\ref{fig:ang_dep_frac_pol}) when we consider the calibration uncertainties mentioned in their paper. \cite{Canovas13} also obtained $H$ and $K_s$ NACO/VLT imaging polarimetry of HD142527 and found the same disk structures, but they derived upper limits for the polarization of $p_{\rm disk}\leq 19~\%$ for the wall on the far side (east) and $p_{\rm disk}\leq 15.8~\%$ for the west side from their $H$-band data. This strongly disagrees with our results and those of \cite{Avenhaus14}, and we suspect that the discrepancy is linked to the extraction of the disk intensity, which is a real challenge for disk data taken with older AO systems under suboptimal conditions.

\subsection{Comparison with other protoplanetary disks}
The normalized disk polarization $Q_{\phi,{\rm disk}}/I_{\rm total}$ strongly depends on the disk geometry, especially on whether we can spatially resolve the strongly illuminated disk regions. The wavelength dependence of the scattered light is not so much a function of geometry, but far more depends on the dust scattering properties. We find a red color between $0.7$ and $1.6~\mu$m for the disk in HD142527, and it is quite well established that other protoplanetary disks also show a similar red wavelength dependence. For example, \cite{Monnier19} derived $Q_{\phi,{\rm disk}}/I_{\rm total}= 1.5$~\% for the $J$-band and 2.5~\% for the $H$-band  for HD34700A. \cite{Stolker16} reported  values of 0.35~\% for the $R$-band, 0.55~\% for the $I$ band, and 0.80~\% for the $J$-band for HD135344B. Similar red colors for the polarized intensity are obtained by Tschudi \& Schmid (in prep.) for HD169142. \citet{Mulders13} and \citet{Sissa18} also derived a red color for the scattered-light intensity of the disk in HD100546, and a gray color for $V-H$ was found for TW Hya by \citet{Debes13}. We are not aware of a protoplanetary disk for which $Q_{\phi,{\rm disk}}/I_{\rm total}$ or $I_{{\rm disk}}/I_{\rm total}$ increase toward shorter wavelengths. Thus, there seems to exist a predominant color trend, and the derived HD142527 results represent an accurate measurement of this color gradient for the polarized intensity of a protoplanetary disk. A red color for the reflected light indicates a dust-scattering albedo that increases toward longer wavelengths. This is predicted, for example, by \citep{Min16b} for dust aggregate models, and this type of dust is also used in the model calculation of the DIANA project \citep{Woitke16}.

The measured degree of polarization $p_{\rm disk}$ of the reflected light strongly depends on the scattering angle, but also on the dust properties. The maximum $p_{\rm disk}$ is expected for a scattering angle $\alpha$ near 90$^\circ$ , and for most disks, there are scattering regions in which $\alpha$ is close to 90$^\circ$, either on the far side as in the case of HD142527 (see Fig.~\ref{fig:tau_0_01_surface}), or near the major axis for disks with higher inclination. Thus, the measured maximum polarization ${\rm max}(p_{\rm disk})$ appears to be another useful diagnostic for constraining the dust in protoplanetary disks. For HD142527 we measured ${\rm max}(p_{\rm disk})\approx 30\%$ for the VBB and $\approx 40\%$ for the $H$-band. Measurements for other disks include those of \citet{Perrin09} with ${\rm max}(p_{\rm disk})\approx 55~\%$ for AB Aur based on NICMOS/HST at 2~$\mu$m data. This result was also confirmed with ground-based imaging polarimetry by \cite{Hashimoto11}. Similar levels of polarization were obtained for HD34700A by \cite{Monnier19} with ${\rm max}(p_{\rm disk})\approx 50~\%$ in the $J$-band and even $\approx 60~\%$ in the $H$-band. A much lower value is found by Tschudi \& Schmid (in prep.) for HD169142, who obtained only about ${\rm max}(p_{\rm disk})=23\pm 5~\%$ for the visual $R'$- and $I'$-bands. More estimates for the polarization of the scattered light from protoplanetary disks are reported in the literature, but usually with uncertainties of about a factor $1.5 - 2$. For example, \cite{Silber00} reported  a degree of polarization of up to 50~\% at 1~$\mu$m for GG Tau, and \cite{Tanii12} measured ${\rm      max}(p_{\rm disk})\approx 65~\%$ for UX Tau A in the $H$-band.

Accurate measurements of the degree of polarization for more disks are urgently needed to clarify whether ${\rm max}(p_{\rm disk})$ differs from system to system and if ${\rm max}(p_{\rm disk})$ is systematically lower for shorter wavelengths. Higher IR polarization may suggest a population of small grains or aggregates with small substructures so that scattering in IR is closer to the Rayleigh regime. The light-scattering grains would have to be relatively large, with sizes $>1\mu$m \citep[e.g.,][]{Mulders13}, to produce the red color in the visual, and simultaneously, the grains would have to exhibit a significant amount of porosity to be able to produce a high maximum polarization despite the large particle size.

\section{Conclusion}
We presented high-precision measurements of the polarized intensity, the intensity, the degree of polarization, and the surface brightness for the scattered light from the extended disk around HD142527 for the two wavelengths $735$~nm and 1.625~$\mu$m based on SPHERE/VLT data. The accuracy we achieved is unprecedented for the visual to near-IR color of the polarized flux and degree of polarization for a protoplanetary disk. Measurements of the wavelength dependence for the scattered light provide a key for determining the scattering properties of the dust because additional ambiguities in the interpretation of the measurements can be lifted.

We compared our measurements with a simple scattering model analysis for the wall on the far side of the disk, taking the near side into account as well. Our analysis yields some rough estimates for the single-scattering albedos ($\omega \approx 0.2-0.5$) and the maximum polarization introduced in the scattering ($P_{\rm max} \approx 0.5-0.75$) and confirms the presence of forward-scattering dust particles ($g\approx 0.5-0.75$), as has been suggested by previous studies. Systematic model calculations applied to our measurements or future high-precision measurements of other protoplanetary disks should provide much better determinations, including the wavelength dependence of these dust scattering parameters. 

Moreover, the measurements of HD142527 presented in this work could be improved with dedicated measurements using SPHERE. In particular, measurements of the disk intensity and the degree of polarization in the IR could be improved with reference star calibrations taken during the same night and by using the star-hopping mode that is available for SPHERE for this purpose. The S/N of the data for the visual data could be improved with longer exposure times in combination with a coronagraph, and the IR data quality would be better for observations without the low-wind effect and in pupil-stabilized mode to minimize the effects of the telescope spiders. Clearly, high-quality observations at more wavelengths from the $V$-band to the $K$-band could also be obtained.

Many scattering models for the dust in protoplanetary disks have been proposed that agree reasonably well for poorly measured disks, but usually show significant discrepancies if the color dependence of the scattering intensity and polarization are also considered. The maximum polarization and the red color for HD142527 found in this work are quite similar to other protoplanetary disks and can be interpreted as the presence of large ($>1\mu$m) dust grains for the red color with a significant amount of porosity, which seems to be necessary to produce the high maximum polarization. However, further studies with detailed dust models are needed to explain the scattering parameters derived with our measurements.

\begin{acknowledgements}
SH and HMS acknowledge the financial support by the Swiss National Science Foundation through grant 200020\_162630/1. SPHERE is an instrument designed and built by a consortium consisting of IPAG (Grenoble, France), MPIA (Heidelberg, Germany), LAM (Marseille, France), LESIA (Paris, France), Laboratoire Lagrange (Nice, France), INAF–Osservatorio di Padova (Italy), Observatoire de Genève (Switzerland), ETH Zurich (Switzerland), NOVA (Netherlands), ONERA (France) and ASTRON (Netherlands) in collaboration with ESO. SPHERE was funded by ESO, with additional contributions from CNRS (France), MPIA (Germany), INAF (Italy), FINES (Switzerland) and NOVA (Netherlands). SPHERE also received funding from the European Commission Sixth and Seventh Framework Programmes as part of the Optical Infrared Coordination Network for Astronomy (OPTICON) under grant number RII3-Ct-2004-001566 for FP6 (2004–2008), grant number 226604 for FP7 (2009–2012) and grant number 312430 for FP7 (2013–2016). We also acknowledge financial support from the Programme National de Planétologie (PNP) and the Programme National de Physique Stellaire (PNPS) of CNRS-INSU in France. This work has also been supported by a grant from the French Labex OSUG@2020 (Investissements d’avenir – ANR10 LABX56). The project is supported by CNRS, by the Agence Nationale de la Recherche (ANR-14-CE33-0018). It has also been carried out within the frame of the National Centre for Competence in Research PlanetS supported by the Swiss National Science Foundation (SNSF). Finally, this work has made use of the the SPHERE Data Centre, jointly operated by OSUG/IPAG (Grenoble), PYTHEAS/LAM/CESAM (Marseille), OCA/Lagrange (Nice), Observatoire de Paris/LESIA (Paris), and Observatoire de Lyon, also supported by a grant from Labex  OSUG@2020 (Investissements d’avenir – ANR10 LABX56). This research has made use of the SIMBAD database, operated at CDS, Strasbourg, France.
\end{acknowledgements}

\bibliographystyle{aa}
\bibliography{paper_ZIMPOL_HD142527_fracpol_sh01}

\begin{thebibliography}{61}
\expandafter\ifx\csname natexlab\endcsname\relax\def\natexlab#1{#1}\fi

\bibitem[{{Avenhaus} {et~al.}(2018){Avenhaus}, {Quanz}, {Garufi}, {Perez},
  {Casassus}, {Pinte}, {Bertrang}, {Caceres}, {Benisty}, \&
  {Dominik}}]{Avenhaus18}
{Avenhaus}, H., {Quanz}, S.~P., {Garufi}, A., {et~al.} 2018, \apj, 863, 44

\bibitem[{{Avenhaus} {et~al.}(2017){Avenhaus}, {Quanz}, {Schmid}, {Dominik},
  {Stolker}, {Ginski}, {de Boer}, {Szul{\'a}gyi}, {Garufi}, {Zurlo},
  {Hagelberg}, {Benisty}, {Henning}, {M{\'e}nard}, {Meyer}, {Baruffolo},
  {Bazzon}, {Beuzit}, {Costille}, {Dohlen}, {Girard}, {Gisler}, {Kasper},
  {Mouillet}, {Pragt}, {Roelfsema}, {Salasnich}, \& {Sauvage}}]{Avenhaus17}
{Avenhaus}, H., {Quanz}, S.~P., {Schmid}, H.~M., {et~al.} 2017, \aj, 154, 33

\bibitem[{{Avenhaus} {et~al.}(2014){Avenhaus}, {Quanz}, {Schmid}, {Meyer},
  {Garufi}, {Wolf}, \& {Dominik}}]{Avenhaus14}
{Avenhaus}, H., {Quanz}, S.~P., {Schmid}, H.~M., {et~al.} 2014, \apj, 781, 87

\bibitem[{{Beuzit} {et~al.}(2019){Beuzit}, {Vigan}, {Mouillet}, {Dohlen},
  {Gratton}, {Boccaletti}, {Sauvage}, {Schmid}, {Langlois}, {Petit},
  {Baruffolo}, {Feldt}, {Milli}, {Wahhaj}, {Abe}, {Anselmi}, {Antichi},
  {Barette}, {Baudrand}, {Baudoz}, {Bazzon}, {Bernardi}, {Blanchard}, {Brast},
  {Bruno}, {Buey}, {Carbillet}, {Carle}, {Cascone}, {Chapron}, {Charton},
  {Chauvin}, {Claudi}, {Costille}, {De Caprio}, {de Boer}, {Delboulb{\'e}},
  {Desidera}, {Dominik}, {Downing}, {Dupuis}, {Fabron}, {Fantinel}, {Farisato},
  {Feautrier}, {Fedrigo}, {Fusco}, {Gigan}, {Ginski}, {Girard}, {Giro},
  {Gisler}, {Gluck}, {Gry}, {Henning}, {Hubin}, {Hugot}, {Incorvaia}, {Jaquet},
  {Kasper}, {Lagadec}, {Lagrange}, {Le Coroller}, {Le Mignant}, {Le Ruyet},
  {Lessio}, {Lizon}, {Llored}, {Lundin}, {Madec}, {Magnard}, {Marteaud},
  {Martinez}, {Maurel}, {M{\'e}nard}, {Mesa}, {M{\"o}ller-Nilsson}, {Moulin},
  {Moutou}, {Orign{\'e}}, {Parisot}, {Pavlov}, {Perret}, {Pragt}, {Puget},
  {Rabou}, {Ramos}, {Reess}, {Rigal}, {Rochat}, {Roelfsema}, {Rousset}, {Roux},
  {Saisse}, {Salasnich}, {Santambrogio}, {Scuderi}, {Segransan}, {Sevin},
  {Siebenmorgen}, {Soenke}, {Stadler}, {Suarez}, {Tiph{\`e}ne}, {Turatto},
  {Udry}, {Vakili}, {Waters}, {Weber}, {Wildi}, {Zins}, \& {Zurlo}}]{Beuzit19}
{Beuzit}, J.~L., {Vigan}, A., {Mouillet}, D., {et~al.} 2019, \aap, 631, A155

\bibitem[{{Biller} {et~al.}(2012){Biller}, {Lacour}, {Juh{\'a}sz}, {Benisty},
  {Chauvin}, {Olofsson}, {Pott}, {M{\"u}ller}, {Sicilia-Aguilar}, {Bonnefoy},
  {Tuthill}, {Thebault}, {Henning}, \& {Crida}}]{Biller12}
{Biller}, B., {Lacour}, S., {Juh{\'a}sz}, A., {et~al.} 2012, \apjl, 753, L38

\bibitem[{{Canovas} {et~al.}(2013){Canovas}, {M{\'e}nard}, {Hales},
  {Jord{\'a}n}, {Schreiber}, {Casassus}, {Gledhill}, \& {Pinte}}]{Canovas13}
{Canovas}, H., {M{\'e}nard}, F., {Hales}, A., {et~al.} 2013, \aap, 556, A123

\bibitem[{{Canovas} {et~al.}(2014){Canovas}, {M{\'e}nard}, {Hales},
  {Jord{\'a}n}, {Schreiber}, {Casassus}, {Gledhill}, \& {Pinte}}]{Canovas14}
{Canovas}, H., {M{\'e}nard}, F., {Hales}, A., {et~al.} 2014, in Revista
  Mexicana de Astronomia y Astrofisica Conference Series, Vol.~44, Revista
  Mexicana de Astronomia y Astrofisica Conference Series, 4--4

\bibitem[{{Cantalloube} {et~al.}(2019){Cantalloube}, {Dohlen}, {Milli},
  {Brandner}, \& {Vigan}}]{Cantalloube19}
{Cantalloube}, F., {Dohlen}, K., {Milli}, J., {Brandner}, W., \& {Vigan}, A.
  2019, The Messenger, 176, 25

\bibitem[{{Casassus} {et~al.}(2013){Casassus}, {Hales}, {de Gregorio}, {Dent},
  {Belloche}, {G{\"u}sten}, {M{\'e}nard}, {Hughes}, {Wilner}, \&
  {Salinas}}]{Casassus13}
{Casassus}, S., {Hales}, A., {de Gregorio}, I., {et~al.} 2013, \aap, 553, A64

\bibitem[{{Casassus} {et~al.}(2015){Casassus}, {Marino}, {P{\'e}rez}, {Roman},
  {Dunhill}, {Armitage}, {Cuadra}, {Wootten}, {van der Plas}, {Cieza}, {Moral},
  {Christiaens}, \& {Montesinos}}]{Casassus15}
{Casassus}, S., {Marino}, S., {P{\'e}rez}, S., {et~al.} 2015, \apj, 811, 92

\bibitem[{{Claudi} {et~al.}(2019){Claudi}, {Maire}, {Mesa}, {Cheetham},
  {Fontanive}, {Gratton}, {Zurlo}, {Avenhaus}, {Bhowmik}, {Biller},
  {Boccaletti}, {Bonavita}, {Bonnefoy}, {Cascone}, {Chauvin}, {Delboulb{\'e}},
  {Desidera}, {D'Orazi}, {Feautrier}, {Feldt}, {Flammini Dotti}, {Girard},
  {Giro}, {Janson}, {Hagelberg}, {Keppler}, {Kopytova}, {Lacour}, {Lagrange},
  {Langlois}, {Lannier}, {Le Coroller}, {Menard}, {Messina}, {Meyer},
  {Millward}, {Olofsson}, {Pavlov}, {Peretti}, {Perrot}, {Pinte}, {Pragt},
  {Ramos}, {Rochat}, {Rodet}, {Roelfsema}, {Rouan}, {Salter}, {Schmidt},
  {Sissa}, {Thebault}, {Udry}, \& {Vigan}}]{Claudi19}
{Claudi}, R., {Maire}, A.~L., {Mesa}, D., {et~al.} 2019, \aap, 622, A96

\bibitem[{{Comer{\'o}n}(2008)}]{Comeron08}
{Comer{\'o}n}, F. 2008, {The Lupus Clouds}, ed. B.~{Reipurth}, Vol.~5, 295

\bibitem[{{Cutri} {et~al.}(2003){Cutri}, {Skrutskie}, {van Dyk}, {Beichman},
  {Carpenter}, {Chester}, {Cambresy}, {Evans}, {Fowler}, {Gizis}, {Howard},
  {Huchra}, {Jarrett}, {Kopan}, {Kirkpatrick}, {Light}, {Marsh}, {McCallon},
  {Schneider}, {Stiening}, {Sykes}, {Weinberg}, {Wheaton}, {Wheelock}, \&
  {Zacarias}}]{Cutri03}
{Cutri}, R.~M., {Skrutskie}, M.~F., {van Dyk}, S., {et~al.} 2003, {2MASS All
  Sky Catalog of point sources.}

\bibitem[{{de Boer} {et~al.}(2020){de Boer}, {Langlois}, {van Holstein},
  {Girard}, {Mouillet}, {Vigan}, {Dohlen}, {Snik}, {Keller}, {Ginski}, {Stam},
  {Milli}, {Wahhaj}, {Kasper}, {Schmid}, {Rabou}, {Gluck}, {Hugot}, {Perret},
  {Martinez}, {Weber}, {Pragt}, {Sauvage}, {Boccaletti}, {Le Coroller},
  {Dominik}, {Henning}, {Lagadec}, {M{\'e}nard}, {Turatto}, {Udry}, {Chauvin},
  {Feldt}, \& {Beuzit}}]{deBoer20}
{de Boer}, J., {Langlois}, M., {van Holstein}, R.~G., {et~al.} 2020, \aap, 633,
  A63

\bibitem[{{Debes} {et~al.}(2013){Debes}, {Jang-Condell}, {Weinberger},
  {Roberge}, \& {Schneider}}]{Debes13}
{Debes}, J.~H., {Jang-Condell}, H., {Weinberger}, A.~J., {Roberge}, A., \&
  {Schneider}, G. 2013, \apj, 771, 45

\bibitem[{{Dohlen} {et~al.}(2008){Dohlen}, {Langlois}, {Saisse}, {Hill},
  {Origne}, {Jacquet}, {Fabron}, {Blanc}, {Llored}, {Carle}, {Moutou}, {Vigan},
  {Boccaletti}, {Carbillet}, {Mouillet}, \& {Beuzit}}]{Dohlen08}
{Dohlen}, K., {Langlois}, M., {Saisse}, M., {et~al.} 2008, in Society of
  Photo-Optical Instrumentation Engineers (SPIE) Conference Series, Vol. 7014,
  Ground-based and Airborne Instrumentation for Astronomy II, 70143L

\bibitem[{{Dominik} {et~al.}(2003){Dominik}, {Dullemond}, {Waters}, \&
  {Walch}}]{Dominik03}
{Dominik}, C., {Dullemond}, C.~P., {Waters}, L.~B.~F.~M., \& {Walch}, S. 2003,
  \aap, 398, 607

\bibitem[{{Duch{\^e}ne} {et~al.}(2004){Duch{\^e}ne}, {McCabe}, {Ghez}, \&
  {Macintosh}}]{Duchene04}
{Duch{\^e}ne}, G., {McCabe}, C., {Ghez}, A.~M., \& {Macintosh}, B.~A. 2004,
  \apj, 606, 969

\bibitem[{{Fujiwara} {et~al.}(2006){Fujiwara}, {Honda}, {Kataza}, {Yamashita},
  {Onaka}, {Fukagawa}, {Okamoto}, {Miyata}, {Sako}, {Fujiyoshi}, \&
  {Sakon}}]{Fujiwara06}
{Fujiwara}, H., {Honda}, M., {Kataza}, H., {et~al.} 2006, \apjl, 644, L133

\bibitem[{{Fukagawa} {et~al.}(2006){Fukagawa}, {Tamura}, {Itoh}, {Kudo},
  {Imaeda}, {Oasa}, {Hayashi}, \& {Hayashi}}]{Fukagawa06}
{Fukagawa}, M., {Tamura}, M., {Itoh}, Y., {et~al.} 2006, \apjl, 636, L153

\bibitem[{{Fukagawa} {et~al.}(2013){Fukagawa}, {Tsukagoshi}, {Momose}, {Saigo},
  {Ohashi}, {Kitamura}, {Inutsuka}, {Muto}, {Nomura}, {Takeuchi}, {Kobayashi},
  {Hanawa}, {Akiyama}, {Honda}, {Fujiwara}, {Kataoka}, {Takahashi}, \&
  {Shibai}}]{Fukagawa13}
{Fukagawa}, M., {Tsukagoshi}, T., {Momose}, M., {et~al.} 2013, \pasj, 65, L14

\bibitem[{{Gaia Collaboration}(2018)}]{GAIA18}
{Gaia Collaboration}. 2018, VizieR Online Data Catalog, I/345

\bibitem[{{Garufi} {et~al.}(2017){Garufi}, {Benisty}, {Stolker}, {Avenhaus},
  {de Boer}, {Pohl}, {Quanz}, {Dominik}, {Ginski}, {Thalmann}, {van Boekel},
  {Boccaletti}, {Henning}, {Janson}, {Salter}, {Schmid}, {Sissa}, {Langlois},
  {Beuzit}, {Chauvin}, {Mouillet}, {Augereau}, {Bazzon}, {Biller}, {Bonnefoy},
  {Buenzli}, {Cheetham}, {Daemgen}, {Desidera}, {Engler}, {Feldt}, {Girard},
  {Gratton}, {Hagelberg}, {Keller}, {Keppler}, {Kenworthy}, {Kral}, {Lopez},
  {Maire}, {Menard}, {Mesa}, {Messina}, {Meyer}, {Milli}, {Min}, {Muller},
  {Olofsson}, {Pawellek}, {Pinte}, {Szulagyi}, {Vigan}, {Wahhaj}, {Waters}, \&
  {Zurlo}}]{Garufi17}
{Garufi}, A., {Benisty}, M., {Stolker}, T., {et~al.} 2017, The Messenger, 169,
  32

\bibitem[{{Gontcharov} \& {Mosenkov}(2019)}]{Gontcharov19}
{Gontcharov}, G.~A. \& {Mosenkov}, A.~V. 2019, \mnras, 483, 299

\bibitem[{{Hashimoto} {et~al.}(2011){Hashimoto}, {Tamura}, {Muto}, {Kudo},
  {Fukagawa}, {Fukue}, {Goto}, {Grady}, {Henning}, {Hodapp}, {Honda},
  {Inutsuka}, {Kokubo}, {Knapp}, {McElwain}, {Momose}, {Ohashi}, {Okamoto},
  {Takami}, {Turner}, {Wisniewski}, {Janson}, {Abe}, {Brandner}, {Carson},
  {Egner}, {Feldt}, {Golota}, {Guyon}, {Hayano}, {Hayashi}, {Hayashi}, {Ishii},
  {Kandori}, {Kusakabe}, {Matsuo}, {Mayama}, {Miyama}, {Morino}, {Moro-Martin},
  {Nishimura}, {Pyo}, {Suto}, {Suzuki}, {Takato}, {Terada}, {Thalmann},
  {Tomono}, {Watanabe}, {Yamada}, {Takami}, \& {Usuda}}]{Hashimoto11}
{Hashimoto}, J., {Tamura}, M., {Muto}, T., {et~al.} 2011, \apjl, 729, L17

\bibitem[{{Hodapp} {et~al.}(2008){Hodapp}, {Suzuki}, {Tamura}, {Abe}, {Suto},
  {Kandori}, {Morino}, {Nishimura}, {Takami}, {Guyon}, {Jacobson},
  {Stahlberger}, {Yamada}, {Shelton}, {Hashimoto}, {Tavrov}, {Nishikawa},
  {Ukita}, {Izumiura}, {Hayashi}, {Nakajima}, {Yamada}, \& {Usuda}}]{Hodapp08}
{Hodapp}, K.~W., {Suzuki}, R., {Tamura}, M., {et~al.} 2008, Society of
  Photo-Optical Instrumentation Engineers (SPIE) Conference Series, Vol. 7014,
  {HiCIAO: the Subaru Telescope's new high-contrast coronographic imager for
  adaptive optics}, 701419

\bibitem[{{H{\o}g} {et~al.}(2000){H{\o}g}, {Fabricius}, {Makarov}, {Urban},
  {Corbin}, {Wycoff}, {Bastian}, {Schwekendiek}, \& {Wicenec}}]{Hog00}
{H{\o}g}, E., {Fabricius}, C., {Makarov}, V.~V., {et~al.} 2000, \aap, 355, L27

\bibitem[{{Honda} {et~al.}(2009){Honda}, {Inoue}, {Fukagawa}, {Oka},
  {Nakamoto}, {Ishii}, {Terada}, {Takato}, {Kawakita}, {Okamoto}, {Shibai},
  {Tamura}, {Kudo}, \& {Itoh}}]{Honda09}
{Honda}, M., {Inoue}, A.~K., {Fukagawa}, M., {et~al.} 2009, \apjl, 690, L110

\bibitem[{{Hunziker} {et~al.}(2020){Hunziker}, {Schmid}, {Mouillet}, {Milli},
  {Zurlo}, {Delorme}, {Abe}, {Avenhaus}, {Baruffolo}, {Bazzon}, {Boccaletti},
  {Baudoz}, {Beuzit}, {Carbillet}, {Chauvin}, {Claudi}, {Costille}, {Daban},
  {Desidera}, {Dohlen}, {Dominik}, {Downing}, {Engler}, {Feldt}, {Fusco},
  {Ginski}, {Gisler}, {Girard}, {Gratton}, {Henning}, {Hubin}, {Kasper},
  {Keller}, {Langlois}, {Lagadec}, {Martinez}, {Maire}, {Menard}, {Meyer},
  {Pavlov}, {Pragt}, {Puget}, {Quanz}, {Rickman}, {Roelfsema}, {Salasnich},
  {Sauvage}, {Siebenmorgen}, {Sissa}, {Snik}, {Suarez}, {Szul{\'a}gyi},
  {Thalmann}, {Turatto}, {Udry}, {van Holstein}, {Vigan}, \&
  {Wildi}}]{Hunziker20}
{Hunziker}, S., {Schmid}, H.~M., {Mouillet}, D., {et~al.} 2020, \aap, 634, A69

\bibitem[{{Malfait} {et~al.}(1998){Malfait}, {Bogaert}, \&
  {Waelkens}}]{Malfait98}
{Malfait}, K., {Bogaert}, E., \& {Waelkens}, C. 1998, \aap, 331, 211

\bibitem[{{Malfait} {et~al.}(1999){Malfait}, {Waelkens}, {Bouwman}, {de Koter},
  \& {Waters}}]{Malfait99}
{Malfait}, K., {Waelkens}, C., {Bouwman}, J., {de Koter}, A., \& {Waters},
  L.~B.~F.~M. 1999, \aap, 345, 181

\bibitem[{{Marino} {et~al.}(2015){Marino}, {Perez}, \& {Casassus}}]{Marino15}
{Marino}, S., {Perez}, S., \& {Casassus}, S. 2015, \apjl, 798, L44

\bibitem[{{Min} {et~al.}(2016{\natexlab{a}}){Min}, {Bouwman}, {Dominik},
  {Waters}, {Pontoppidan}, {Hony}, {Mulders}, {Henning}, {van Dishoeck},
  {Woitke}, {Evans}, \& {Digit Team}}]{Min16}
{Min}, M., {Bouwman}, J., {Dominik}, C., {et~al.} 2016{\natexlab{a}}, \aap,
  593, A11

\bibitem[{{Min} {et~al.}(2016{\natexlab{b}}){Min}, {Rab}, {Woitke}, {Dominik},
  \& {M{\'e}nard}}]{Min16b}
{Min}, M., {Rab}, C., {Woitke}, P., {Dominik}, C., \& {M{\'e}nard}, F.
  2016{\natexlab{b}}, \aap, 585, A13

\bibitem[{{Monnier} {et~al.}(2017){Monnier}, {Harries}, {Aarnio}, {Adams},
  {Andrews}, {Calvet}, {Espaillat}, {Hartmann}, {Hinkley}, {Kraus}, {McClure},
  {Oppenheimer}, {Perrin}, \& {Wilner}}]{Monnier17}
{Monnier}, J.~D., {Harries}, T.~J., {Aarnio}, A., {et~al.} 2017, \apj, 838, 20

\bibitem[{{Monnier} {et~al.}(2019){Monnier}, {Harries}, {Bae}, {Setterholm},
  {Laws}, {Aarnio}, {Adams}, {Andrews}, {Calvet}, {Espaillat}, {Hartmann},
  {Kraus}, {McClure}, {Miller}, {Oppenheimer}, {Wilner}, \& {Zhu}}]{Monnier19}
{Monnier}, J.~D., {Harries}, T.~J., {Bae}, J., {et~al.} 2019, \apj, 872, 122

\bibitem[{{Mulders} {et~al.}(2013){Mulders}, {Min}, {Dominik}, {Debes}, \&
  {Schneider}}]{Mulders13}
{Mulders}, G.~D., {Min}, M., {Dominik}, C., {Debes}, J.~H., \& {Schneider}, G.
  2013, \aap, 549, A112

\bibitem[{{Pavlov} {et~al.}(2008){Pavlov}, {M{\"o}ller-Nilsson}, {Feldt},
  {Henning}, {Beuzit}, \& {Mouillet}}]{Pavlov08}
{Pavlov}, A., {M{\"o}ller-Nilsson}, O., {Feldt}, M., {et~al.} 2008, in
  \procspie, Vol. 7019, Advanced Software and Control for Astronomy II, 701939

\bibitem[{{Perez} {et~al.}(2015){Perez}, {Casassus}, {M{\'e}nard}, {Roman},
  {van der Plas}, {Cieza}, {Pinte}, {Christiaens}, \& {Hales}}]{Perez15}
{Perez}, S., {Casassus}, S., {M{\'e}nard}, F., {et~al.} 2015, \apj, 798, 85

\bibitem[{{Perrin} {et~al.}(2015){Perrin}, {Duchene}, {Millar-Blanchaer},
  {Fitzgerald}, {Graham}, {Wiktorowicz}, {Kalas}, {Macintosh}, {Bauman},
  {Cardwell}, {Chilcote}, {De Rosa}, {Dillon}, {Doyon}, {Dunn}, {Erikson},
  {Gavel}, {Goodsell}, {Hartung}, {Hibon}, {Ingraham}, {Kerley}, {Konapacky},
  {Larkin}, {Maire}, {Marchis}, {Marois}, {Mittal}, {Morzinski}, {Oppenheimer},
  {Palmer}, {Patience}, {Poyneer}, {Pueyo}, {Rantakyr{\"o}}, {Sadakuni},
  {Saddlemyer}, {Savransky}, {Soummer}, {Sivaramakrishnan}, {Song}, {Thomas},
  {Wallace}, {Wang}, \& {Wolff}}]{Perrin15}
{Perrin}, M.~D., {Duchene}, G., {Millar-Blanchaer}, M., {et~al.} 2015, \apj,
  799, 182

\bibitem[{{Perrin} {et~al.}(2009){Perrin}, {Schneider}, {Duchene}, {Pinte},
  {Grady}, {Wisniewski}, \& {Hines}}]{Perrin09}
{Perrin}, M.~D., {Schneider}, G., {Duchene}, G., {et~al.} 2009, \apjl, 707,
  L132

\bibitem[{{Pinte} {et~al.}(2008){Pinte}, {Padgett}, {M{\'e}nard},
  {Stapelfeldt}, {Schneider}, {Olofsson}, {Pani{\'c}}, {Augereau},
  {Duch{\^e}ne}, {Krist}, {Pontoppidan}, {Perrin}, {Grady}, {Kessler-Silacci},
  {van Dishoeck}, {Lommen}, {Silverstone}, {Hines}, {Wolf}, {Blake}, {Henning},
  \& {Stecklum}}]{Pinte08}
{Pinte}, C., {Padgett}, D.~L., {M{\'e}nard}, F., {et~al.} 2008, \aap, 489, 633

\bibitem[{{Pottasch} \& {Parthasarathy}(1988)}]{Pottasch88}
{Pottasch}, S.~R. \& {Parthasarathy}, M. 1988, \aap, 192, 182

\bibitem[{{Price} {et~al.}(2018){Price}, {Cuello}, {Pinte}, {Mentiplay},
  {Casassus}, {Christiaens}, {Kennedy}, {Cuadra}, {Sebastian Perez}, {Marino},
  {Armitage}, {Zurlo}, {Juhasz}, {Ragusa}, {Laibe}, \& {Lodato}}]{Price18}
{Price}, D.~J., {Cuello}, N., {Pinte}, C., {et~al.} 2018, \mnras, 477, 1270

\bibitem[{{Quanz} {et~al.}(2011){Quanz}, {Schmid}, {Geissler}, {Meyer},
  {Henning}, {Brandner}, \& {Wolf}}]{Quanz11}
{Quanz}, S.~P., {Schmid}, H.~M., {Geissler}, K., {et~al.} 2011, \apj, 738, 23

\bibitem[{{Rizzo} {et~al.}(1998){Rizzo}, {Morras}, \& {Arnal}}]{Rizzo98}
{Rizzo}, J.~R., {Morras}, R., \& {Arnal}, E.~M. 1998, \mnras, 300, 497

\bibitem[{{Roddier} {et~al.}(1995){Roddier}, {Roddier}, {Graves}, \&
  {Northcott}}]{Roddier95}
{Roddier}, F., {Roddier}, C., {Graves}, J.~E., \& {Northcott}, M.~J. 1995,
  \apj, 443, 249

\bibitem[{{Rosenfeld} {et~al.}(2014){Rosenfeld}, {Chiang}, \&
  {Andrews}}]{Rosenfeld14}
{Rosenfeld}, K.~A., {Chiang}, E., \& {Andrews}, S.~M. 2014, \apj, 782, 62

\bibitem[{{Sauvage} {et~al.}(2015){Sauvage}, {Fusco}, {Guesalaga},
  {Wizinowitch}, {O'Neal}, {N'Diaye}, {Vigan}, {Girard}, {Lesur}, {Mouillet},
  {Buezit}, {Kasper}, {Le Louarn}, {Mlli}, {Dohlen}, {Neichel}, {Bourget},
  {Heigenauer}, \& {Mawet}}]{Sauvage15}
{Sauvage}, J.-F., {Fusco}, T., {Guesalaga}, A., {et~al.} 2015, in Adaptive
  Optics for Extremely Large Telescopes IV (AO4ELT4), E9

\bibitem[{{Schmid} {et~al.}(2018){Schmid}, {Bazzon}, {Roelfsema}, {Mouillet},
  {Milli}, {Menard}, {Gisler}, {Hunziker}, {Pragt}, {Dominik}, {Boccaletti},
  {Ginski}, {Abe}, {Antoniucci}, {Avenhaus}, {Baruffolo}, {Baudoz}, {Beuzit},
  {Carbillet}, {Chauvin}, {Claudi}, {Costille}, {Daban}, {de Haan}, {Desidera},
  {Dohlen}, {Downing}, {Elswijk}, {Engler}, {Feldt}, {Fusco}, {Girard},
  {Gratton}, {Hanenburg}, {Henning}, {Hubin}, {Joos}, {Kasper}, {Keller},
  {Langlois}, {Lagadec}, {Martinez}, {Mulder}, {Pavlov}, {Podio}, {Puget},
  {Quanz}, {Rigal}, {Salasnich}, {Sauvage}, {Schuil}, {Siebenmorgen}, {Sissa},
  {Snik}, {Suarez}, {Thalmann}, {Turatto}, {Udry}, {van Duin}, {van Holstein},
  {Vigan}, \& {Wildi}}]{Schmid18}
{Schmid}, H.~M., {Bazzon}, A., {Roelfsema}, R., {et~al.} 2018, \aap, 619, A9

\bibitem[{{Schmid} {et~al.}(2006){Schmid}, {Joos}, \& {Tschan}}]{Schmid06b}
{Schmid}, H.~M., {Joos}, F., \& {Tschan}, D. 2006, \aap, 452, 657

\bibitem[{{Serkowski} {et~al.}(1975){Serkowski}, {Mathewson}, \&
  {Ford}}]{Serkowski75}
{Serkowski}, K., {Mathewson}, D.~S., \& {Ford}, V.~L. 1975, \apj, 196, 261

\bibitem[{{Silber} {et~al.}(2000){Silber}, {Gledhill}, {Duch{\^e}ne}, \&
  {M{\'e}nard}}]{Silber00}
{Silber}, J., {Gledhill}, T., {Duch{\^e}ne}, G., \& {M{\'e}nard}, F. 2000,
  \apjl, 536, L89

\bibitem[{{Sissa} {et~al.}(2018){Sissa}, {Gratton}, {Garufi}, {Rigliaco},
  {Zurlo}, {Mesa}, {Langlois}, {de Boer}, {Desidera}, {Ginski}, {Lagrange},
  {Maire}, {Vigan}, {Dima}, {Antichi}, {Baruffolo}, {Bazzon}, {Benisty},
  {Beuzit}, {Biller}, {Boccaletti}, {Bonavita}, {Bonnefoy}, {Brandner},
  {Bruno}, {Buenzli}, {Cascone}, {Chauvin}, {Cheetham}, {Claudi}, {Cudel}, {De
  Caprio}, {Dominik}, {Fantinel}, {Farisato}, {Feldt}, {Fontanive}, {Galicher},
  {Giro}, {Hagelberg}, {Incorvaia}, {Janson}, {Kasper}, {Keppler}, {Kopytova},
  {Lagadec}, {Lannier}, {Lazzoni}, {LeCoroller}, {Lessio}, {Ligi}, {Marzari},
  {Menard}, {Meyer}, {Mouillet}, {Peretti}, {Perrot}, {Potiron}, {Rouan},
  {Salasnich}, {Salter}, {Samland}, {Schmidt}, {Scuderi}, \& {Wildi}}]{Sissa18}
{Sissa}, E., {Gratton}, R., {Garufi}, A., {et~al.} 2018, \aap, 619, A160

\bibitem[{{Stolker} {et~al.}(2016){Stolker}, {Dominik}, {Avenhaus}, {Min}, {de
  Boer}, {Ginski}, {Schmid}, {Juhasz}, {Bazzon}, {Waters}, {Garufi},
  {Augereau}, {Benisty}, {Boccaletti}, {Henning}, {Langlois}, {Maire},
  {M{\'e}nard}, {Meyer}, {Pinte}, {Quanz}, {Thalmann}, {Beuzit}, {Carbillet},
  {Costille}, {Dohlen}, {Feldt}, {Gisler}, {Mouillet}, {Pavlov}, {Perret},
  {Petit}, {Pragt}, {Rochat}, {Roelfsema}, {Salasnich}, {Soenke}, \&
  {Wildi}}]{Stolker16}
{Stolker}, T., {Dominik}, C., {Avenhaus}, H., {et~al.} 2016, \aap, 595, A113

\bibitem[{{Takami} {et~al.}(2014){Takami}, {Hasegawa}, {Muto}, {Gu}, {Dong},
  {Karr}, {Hashimoto}, {Kusakabe}, {Chapillon}, {Tang}, {Itoh}, {Carson},
  {Follette}, {Mayama}, {Sitko}, {Janson}, {Grady}, {Kudo}, {Akiyama}, {Kwon},
  {Takahashi}, {Suenaga}, {Abe}, {Brandner}, {Brand t}, {Currie}, {Egner},
  {Feldt}, {Guyon}, {Hayano}, {Hayashi}, {Hayashi}, {Henning}, {Hodapp},
  {Honda}, {Ishii}, {Iye}, {Kandori}, {Knapp}, {Kuzuhara}, {McElwain},
  {Matsuo}, {Miyama}, {Morino}, {Moro-Martin}, {Nishimura}, {Pyo}, {Serabyn},
  {Suto}, {Suzuki}, {Takato}, {Terada}, {Thalmann}, {Tomono}, {Turner},
  {Wisniewski}, {Watanabe}, {Yamada}, {Takami}, {Usuda}, \&
  {Tamura}}]{Takami14}
{Takami}, M., {Hasegawa}, Y., {Muto}, T., {et~al.} 2014, \apj, 795, 71

\bibitem[{{Tanii} {et~al.}(2012){Tanii}, {Itoh}, {Kudo}, {Hioki}, {Oasa},
  {Gupta}, {Sen}, {Wisniewski}, {Muto}, {Grady}, {Hashimoto}, {Fukagawa},
  {Mayama}, {Hornbeck}, {Sitko}, {Russell}, {Werren}, {Cur{\'e}}, {Currie},
  {Ohashi}, {Okamoto}, {Momose}, {Honda}, {Inutsuka}, {Takeuchi}, {Dong},
  {Abe}, {Brand ner}, {Brandt}, {Carson}, {Egner}, {Feldt}, {Fukue}, {Goto},
  {Guyon}, {Hayano}, {Hayashi}, {Hayashi}, {Henning}, {Hodapp}, {Ishii}, {Iye},
  {Janson}, {Kandori}, {Knapp}, {Kusakabe}, {Kuzuhara}, {Matsuo}, {McElwain},
  {Miyama}, {Morino}, {Moro-Mart{\'\i}n}, {Nishimura}, {Pyo}, {Serabyn},
  {Suto}, {Suzuki}, {Takami}, {Takato}, {Terada}, {Thalmann}, {Tomono},
  {Turner}, {Watanabe}, {Yamada}, {Takami}, {Usuda}, \& {Tamura}}]{Tanii12}
{Tanii}, R., {Itoh}, Y., {Kudo}, T., {et~al.} 2012, \pasj, 64, 124

\bibitem[{{van Holstein} {et~al.}(2020){van Holstein}, {Girard}, {de Boer},
  {Snik}, {Milli}, {Stam}, {Ginski}, {Mouillet}, {Wahhaj}, {Schmid}, {Keller},
  {Langlois}, {Dohlen}, {Vigan}, {Pohl}, {Carbillet}, {Fantinel}, {Maurel},
  {Orign{\'e}}, {Petit}, {Ramos}, {Rigal}, {Sevin}, {Boccaletti}, {Le
  Coroller}, {Dominik}, {Henning}, {Lagadec}, {M{\'e}nard}, {Turatto}, {Udry},
  {Chauvin}, {Feldt}, \& {Beuzit}}]{vanHolstein20}
{van Holstein}, R.~G., {Girard}, J.~H., {de Boer}, J., {et~al.} 2020, \aap,
  633, A64

\bibitem[{{van Holstein} {et~al.}(2017){van Holstein}, {Snik}, {Girard}, {de
  Boer}, {Ginski}, {Keller}, {Stam}, {Beuzit}, {Mouillet}, {Kasper},
  {Langlois}, {Zurlo}, {de Kok}, \& {Vigan}}]{vanHolstein17}
{van Holstein}, R.~G., {Snik}, F., {Girard}, J.~H., {et~al.} 2017, in Society
  of Photo-Optical Instrumentation Engineers (SPIE) Conference Series, Vol.
  10400, Society of Photo-Optical Instrumentation Engineers (SPIE) Conference
  Series, 1040015

\bibitem[{{Verhoeff} {et~al.}(2011){Verhoeff}, {Min}, {Pantin}, {Waters},
  {Tielens}, {Honda}, {Fujiwara}, {Bouwman}, {van Boekel}, {Dougherty}, {de
  Koter}, {Dominik}, \& {Mulders}}]{Verhoeff11}
{Verhoeff}, A.~P., {Min}, M., {Pantin}, E., {et~al.} 2011, \aap, 528, A91

\bibitem[{{Woitke} {et~al.}(2016){Woitke}, {Min}, {Pinte}, {Thi}, {Kamp},
  {Rab}, {Anthonioz}, {Antonellini}, {Baldovin-Saavedra}, {Carmona}, {Dominik},
  {Dionatos}, {Greaves}, {G{\"u}del}, {Ilee}, {Liebhart}, {M{\'e}nard},
  {Rigon}, {Waters}, {Aresu}, {Meijerink}, \& {Spaans}}]{Woitke16}
{Woitke}, P., {Min}, M., {Pinte}, C., {et~al.} 2016, \aap, 586, A103

\end{thebibliography}

\clearpage

\begin{appendix}
\section{Radial profiles}
\label{sec:Radial profiles}

\subsection{VBB data}
\label{sec:VBB radial profiles}
\begin{figure*}[b]
\centering
\begin{tabular}{cc}
\includegraphics[width=9.0cm]{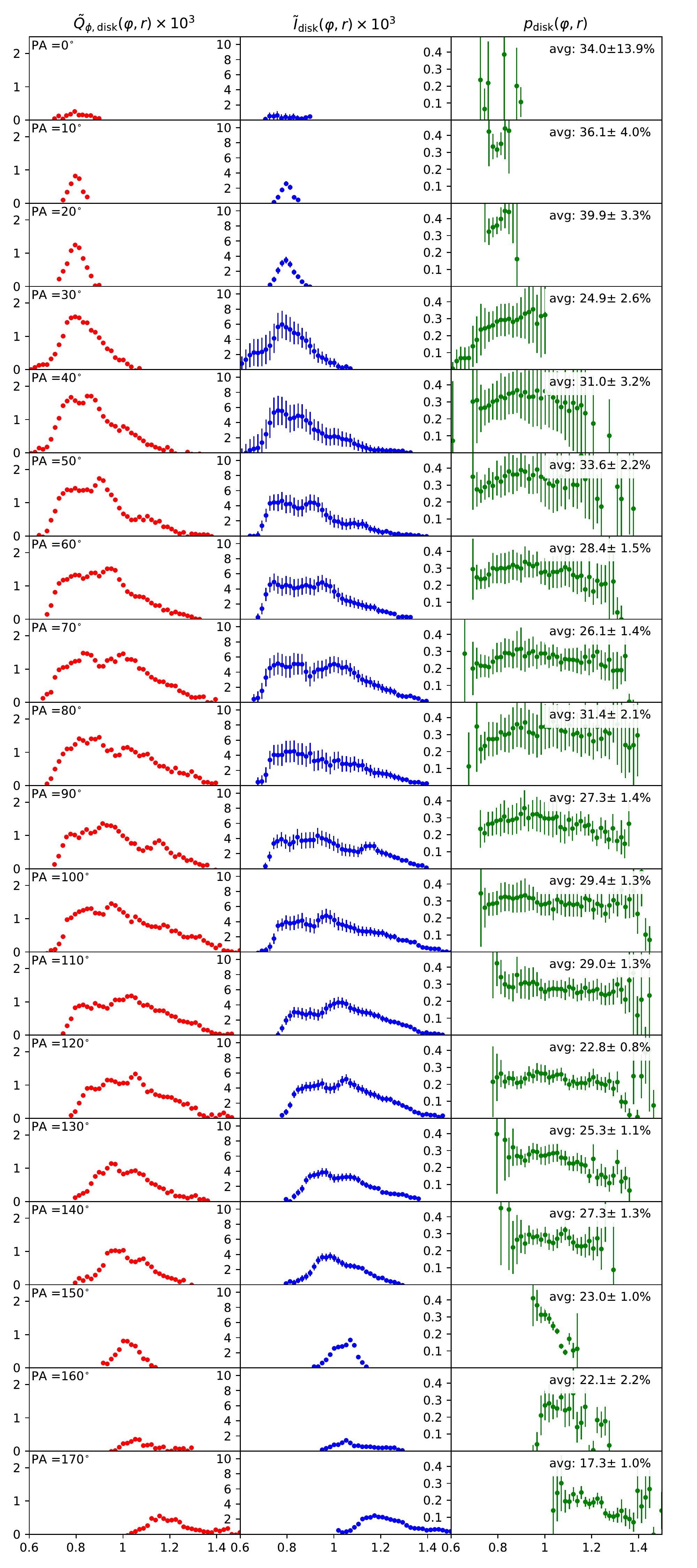}
\includegraphics[width=9.0cm]{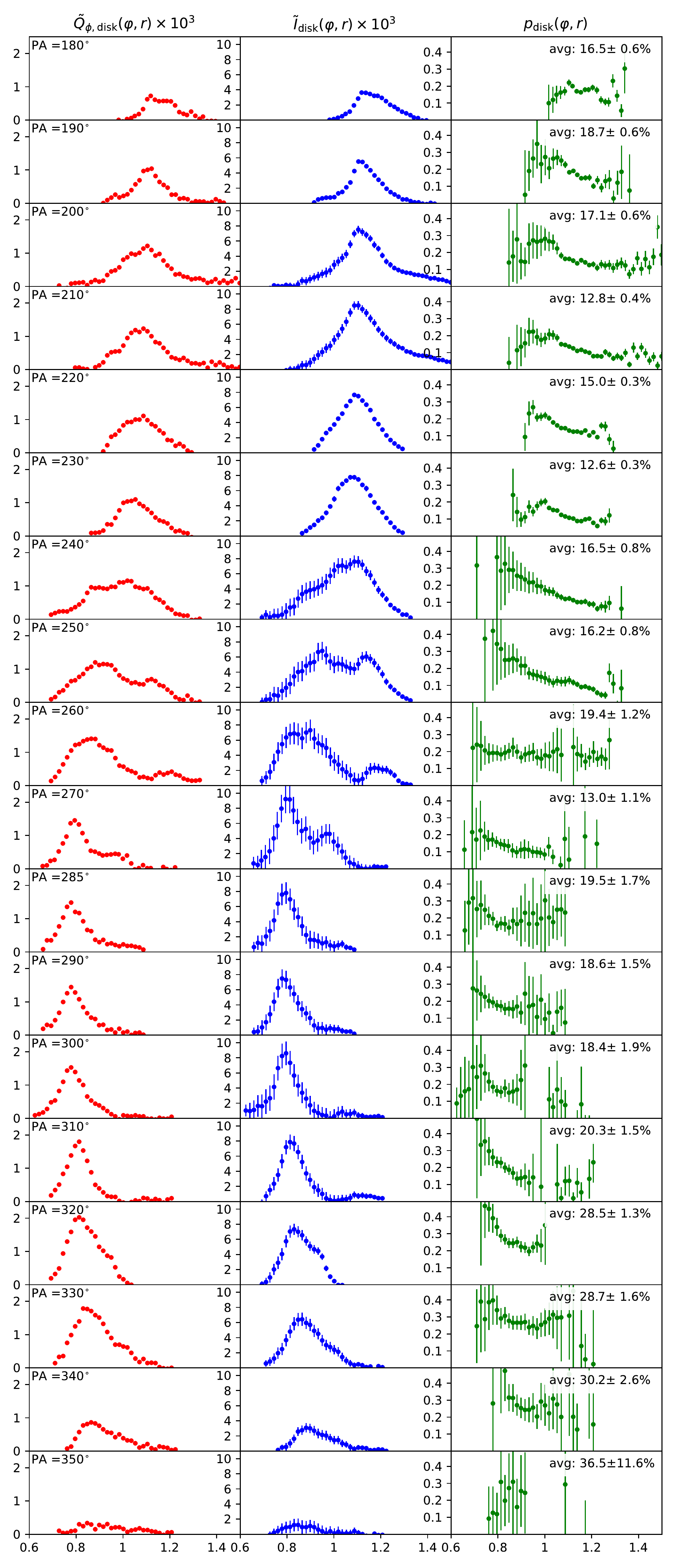} \\
\end{tabular}
\caption{Radial surface brightness profiles for Q$_{\phi}$ (red), Stokes I (blue), and the degree of polarization Q$_{\phi}/$I (green) at several different position angles for the ZIMPOL VBB observations. The intensity values on the vertical axis are in units of surface brightness relative to the total intensity of the system, i.e., [$I_{\rm total}/{\rm arcsec}^2$].}
  \label{fig:HD142527_VBB_radial_profiles}
\end{figure*}

\clearpage

\subsection{$H$-band data}
\label{sec:Hband radial profiles}
\begin{figure*}[b]
\centering
\begin{tabular}{cc}
\includegraphics[width=9.0cm]{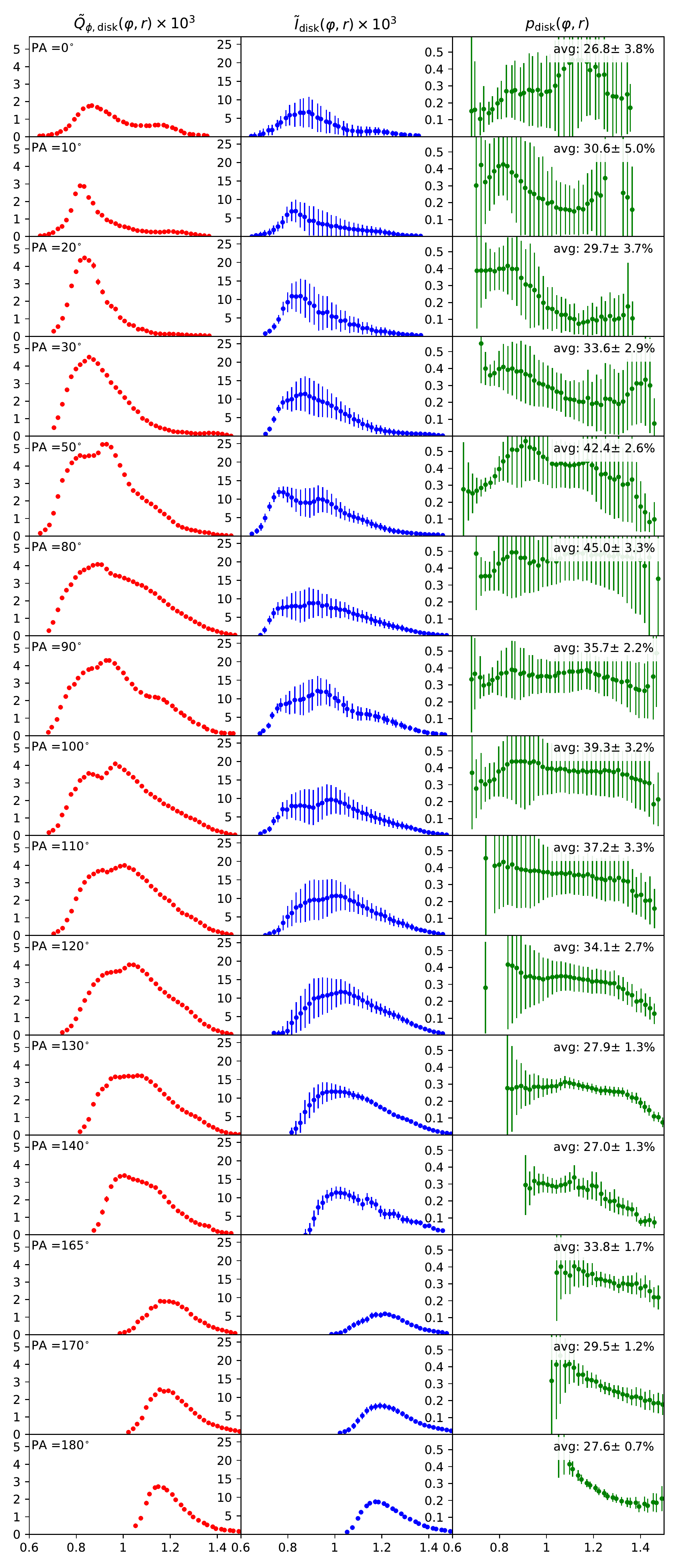}
\includegraphics[width=9.0cm]{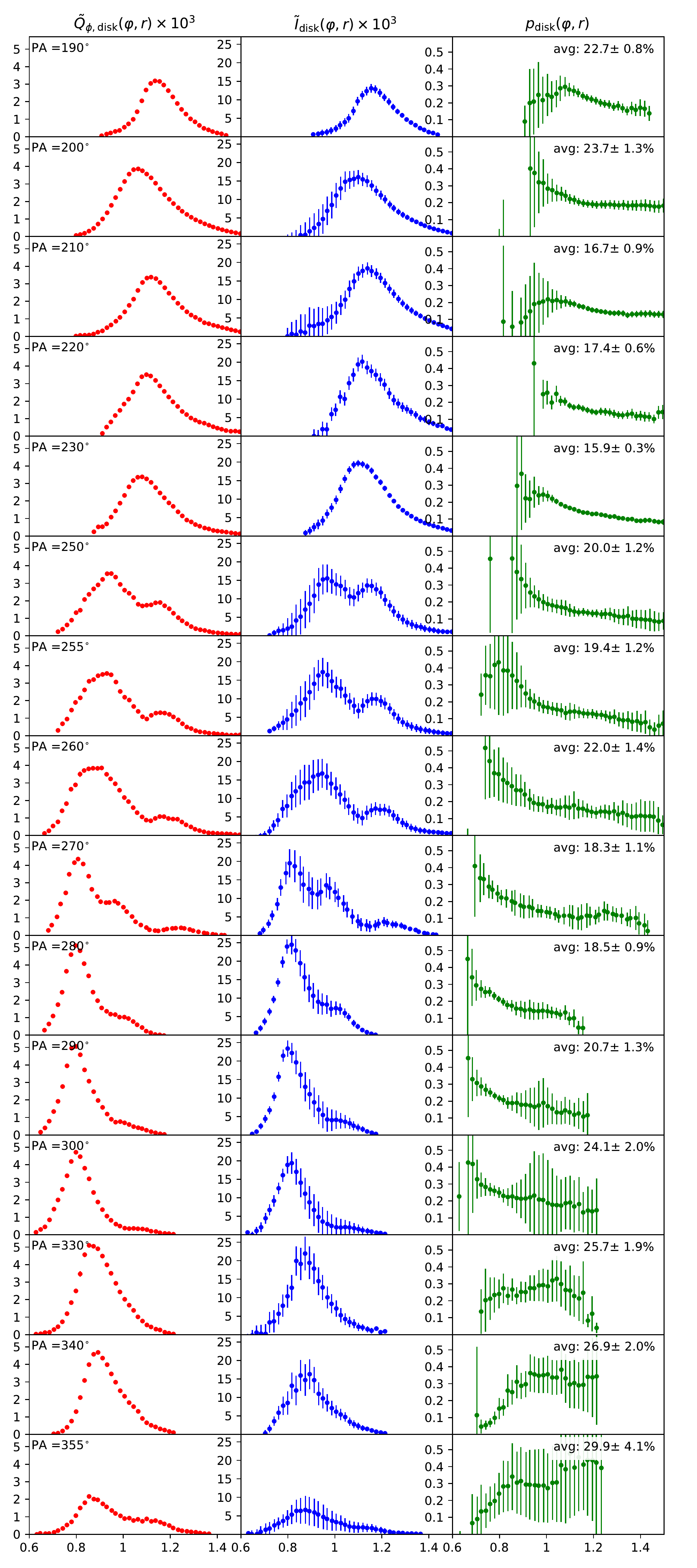} \\
\end{tabular}
\caption{Radial surface brightness profiles for Q$_{\phi}$ (red), Stokes I (blue), and the degree of polarization Q$_{\phi}/$I (green) at several different position angles for the IRDIS $H$-band observations. The intensity values on the vertical axis are in units of surface brightness relative to the total intensity of the system, i.e., [$I_{\rm total}/{\rm arcsec}^2$].}
  \label{fig:HD142527_Hband_radial_profiles}
\end{figure*}

\end{appendix}

\end{document}